\documentclass[12pt,preprint]{aastex}

\slugcomment{Cha~II PMS population: \today}

\shorttitle{The PMS population in Chamaeleon II}
\shortauthors{Spezzi et al.}

\begin{document}


\title{The young population of the Chamaeleon~II dark cloud}



\author{
Loredana~Spezzi\altaffilmark{1,2},  
Juan~M.~Alcal\'a\altaffilmark{1},
Elvira~Covino\altaffilmark{1},
Antonio~Frasca\altaffilmark{2},
Davide~Gandolfi\altaffilmark{2},
Isa~Oliveira\altaffilmark{3,4},
Nicholas~Chapman\altaffilmark{6},
Neal~J.~Evans~II\altaffilmark{7},
Tracy~L.~Huard\altaffilmark{8},
Jes~K.~J{\o}rgensen\altaffilmark{9},
Bruno~Mer\'{i}n\altaffilmark{5,4},
Karl~R.~Stapelfeldt\altaffilmark{10}
}

 

\altaffiltext{1}{INAF-OA Capodimonte, via Moiariello 16, 80131, Naples, Italy; lspezzi@oact.inaf.it}
\altaffiltext{2}{INAF-OA Catania, via S. Sofia, 78, 95123 Catania, Italy; afr@oact.inaf.it}
\altaffiltext{3}{Division of Geological and Planetary Sciences, California Institute of Technology, Pasadena, CA 91125, USA; isa@gps.caltech.edu}
\altaffiltext{4}{Leiden Observatory, Leiden University, P.O. Box 9513, 2300 RA Leiden, The Netherlands}
\altaffiltext{5}{Research and Scientific Support Dept. (ESTEC), European Space Agency, Keplerlaan, 1,  PO Box 299, 2200 AG Noordwijk, The Netherlands; bmerin@rssd.esa.int}
\altaffiltext{6}{Astronomy Department, University of Maryland, College Park, MD~20742; chapman@astro.umd.edu}
\altaffiltext{7}{Astronomy Department, University of Texas at Austin, 1 University Station C1400, Austin, TX~78712-0259; nje@astro.as.utexas.edu}
\altaffiltext{8}{Smithsonian Astrophysical Observatory, 60 Garden Street, MS42, Cambridge, MA~02138; thuard@cfa.harvard.edu;}
\altaffiltext{9}{Argelander-Institut f\"ur Astronomie, University of Bonn, Auf dem H\"ugel 71, 53121 Bonn, Germany; jes@astro.uni-bonn.de}
\altaffiltext{10}{Jet Propulsion Laboratory, California Institute of Technology, Pasadena, CA; krs@exoplanet.jpl.nasa.gov.}


\begin{abstract}

We discuss the results of the optical spectroscopic follow-up of pre-main 
sequence (PMS) objects and candidates selected in the Chamaeleon~II dark
cloud based on data from the Spitzer Legacy survey
\emph{From Molecular Cores to Planet Forming Disks} (c2d) and from
previous surveys. Our sample includes both objects with infrared excess 
selected according to c2d criteria and referred to as Young Stellar 
Objects and other cloud members and candidates selected from 
complementary optical and near-infrared data.


We characterize the sample of objects by deriving their physical parameters.
The vast majority of objects have masses $M \leq 1 M_{\odot}$ and ages $<$ 6 Myr. 
Several of the PMS objects and candidates lie very close to or below the 
Hydrogen-burning limit.
A first estimate of the slope of the Initial Mass Function in Cha~II
is consistent with that of other T associations. The star formation efficiency 
in the cloud (1-4\%) is consistent with our own estimates for Taurus and Lupus,
but significantly lower than for Cha~I. This might mean that different
star-formation activities in the Chamaeleon clouds may reflect a different
history of star formation.
We also find that the Cha~II cloud is turning some 8~$M_{\odot}$ into
stars every Myr, which is less than the star formation rate in the other
c2d clouds. However, the star formation rate is not steady and evidence
is found that the star formation in Cha~II might have occurred very rapidly.
The H$\alpha$ emission of the Cha~II PMS objects, as well as  possible 
correlations between their stellar and disk properties, are also 
investigated.

\end{abstract}





\newpage
\section{Introduction}

During the last decade, many efforts have been devoted to the investigation of
pre-main sequence (PMS) objects in star-forming regions  and young clusters
\citep[][and references therein]{Hil00, Luh00, Bri02, Dow06, Pre03, Mue03, Dow06},
with the aim of studying the star formation process down to the very
low-mass end of the Initial Mass Function (IMF). Unlike field stars and sub-stellar
objects of the solar neighborhood, young populations in star forming regions and 
young clusters allow the interpretation of their observational quantities free from 
caveats of evolutionary or dynamical biases, because their properties are better 
constrained.
The wealth of data collected so far indicates that stars and brown dwarfs (BDs)
share a common formation history. Indeed, the detection of disks around BDs
\citep{Luh05a, Luh06, Jay06, Alc06}, the similarity of the disk
fractions of stars and BDs \citep{Jay03a, Luh05b, Liu03}
and the continuity of accretion rates from stars to BDs
\citep{Muz00, Muz03, Muz05, Bri02, Whi03, Jay03b, Moh05},
suggest a stellar-like formation process,
down to the Deuterium burning limit \citep{Luh07}.
Recent investigations have pointed out, however, that the fraction of young BDs
relative to low-mass and more massive PMS stars may vary significantly among
different star forming regions \citep{Kro02}. In particular, the fraction in OB associations
\citep{Bou98, Hil00, Bar01, Bri02, Mue02} exceeds that determined for T associations
\citep{Com00, Bej01, Tej02, Bri02, Spe07}, suggesting that the environment may have
an important impact on the BD formation process.

The Spitzer Space Telescope offers the opportunity for a major advance in the study
of star and BD formation, overcoming the problems that generally affect the observational
approaches to this subject (i.e. photometric and spatial incompleteness, incomplete
data across the wavelength ranges of interest, etc).
The Spitzer Legacy Survey \emph{From Molecular Cores to Planet Forming Disks} (c2d)
has been completely devoted to studying the process of star and planet formation from the
earliest stages of molecular cores to the epoch of planet-forming disks \citep{Eva03}.
In the context of the c2d survey, five large molecular clouds have been selected
for mapping, with the criterion of encompassing all modes of star formation.
The Chamaeleon~II (Cha~II) dark cloud, at a distance of 178~pc \citep{Whi97},
has been chosen as a test-case of a region with low extinction and modest star-formation
activity, whose young stars are isolated or in sparse groups. Cha~II has been mapped
with the Infrared Array Camera \citep[IRAC,][]{Por07} and the Multiband 
Imaging Photometer for Spitzer \citep[MIPS,][]{You05}; 
in addition, complementary optical imaging 
data have been obtained with the Wide-Field Imager 
(WFI) at the ESO 2.2m telescope \citep{Spe07} and a
deep near-IR survey of the cloud was performed by \citet{All06a}. The results of these
combined observations have been reported in the synthesis work by
\citet[][hereafter Paper~I]{Alc07}, which provides a reliable census of the young
population in Cha~II, down to 0.03~M$_{\odot}$, and investigates the IR properties and
circumstellar disks of the young population through the comparison of the spectral
energy distributions (SEDs) and colors with recent passive and accretion disk models
for young stellar and sub-stellar objects. By merging the c2d data with the collection
of data from optical to millimeter wavelengths available in the literature, Paper~I
also presents an overview on the global properties of Cha~II, such as clustering,
extinction, cloud mass, star formation efficiency, star formation rate, etc.

The main goal of the present paper, is the characterization of the young population in Cha~II
by the determination of the physical parameters of the cloud members. To this aim, we
present optical spectroscopic follow-up observations of the objects selected in Paper~I.
The optical spectroscopy allows us to confirm the young nature of the PMS objects and
candidates reported in Paper~I, as well as to improve the spectral type classification
of the previously known members of the cloud. Once the membership is established, we
determine the effective temperature and luminosity for all the cloud members and hence,
their masses and ages by comparison with theoretical PMS evolutionary tracks.
This procedure allows us to investigate the star formation history in Cha~II and estimate
the IMF of the region. We also discuss the stellar properties and spectroscopic
features and their link with the disk properties determined in Paper~I. This work
is thus intended to be a complementary companion paper to Paper~I.

Throughout this paper we will adopt the same terminology of PMS and
young stellar objects (YSOs) as presented in Paper~I; i.e. we define YSOs
to be those PMS objects with detectable IR excess in at least one of the Spitzer bands, 
that satisfy the c2d multi-color selection criteria described in \citet{Har07}, whereas
the more general term ``PMS object'' refers to young objects, not yet on the main
sequence, that may or may not possess IR excess and which may have been discovered
in  H$\alpha$ or X-ray surveys. In this sense, all YSOs are PMS objects,
but not vice versa.

In \S~\ref{sample} we describe the sample of PMS objects and candidates investigated
in this work and present an overview of the spectroscopic observations and data
reduction. The results of the follow-up spectroscopy are reported in \S~\ref{results}
and \S~\ref{par}, where we also  describe the procedures adopted for the spectral
type classification and the determination of their physical parameters.
In \S~\ref{substellar_objs} we focus on the sub-stellar members of Cha~II, while
in \S~\ref{SF} we report the overall results on the IMF and the star formation
history in the cloud. Finally, in \S~\ref{star_disk} and \S~\ref{Ha_accr_ind} we
discuss possible correlations between spectroscopic diagnostics, stellar
parameters, and disk properties. We summarize our results in \S~\ref{sum}.

\section{The sample and the observations}

\subsection{The sample}
\label{sample}

The sample studied in this work consists of the 62 PMS objects and candidates 
reported in Paper~I. Several of these objects were proposed as potential members 
of Cha~II in previous works \citep{Sch77, Sch91, Gau92, Hug92, Har93, Alc95, Vuo01,
Bar04} based on their strong IR excess emission, H$\alpha$
emission or X-ray emission, but for many of them the confirmation
of the Lithium $\lambda$6707.8~\AA~ absorption line in their optical
spectra was missing.
The presence of strong absorption in the Li{\sc i} resonance line
represents in fact the most important criterion for the identification
of low-mass PMS stars, since lithium is efficiently destroyed by convective
mixing in the stellar interior when the temperature at the bottom of the
convective layer reaches about 2$\times$10$^6$~K \citep{Bod65}.

Likewise, the candidates selected in the surveys
by \citet{You05} and \citet{Spe07} and those resulting
from the c2d selection criteria in Paper~I, as well as the sub-stellar
objects reported by \citet{All06a} and \citet{Alc06} are included.
Table~\ref{pms_objects_a} reports the studied sample. For the sake
of homogeneity the table is presented in the same format as in Paper~I
and, for completeness, we also include the entry of the outflow HH~54,
as well as the three Class~I objects discussed in Paper~I.
As will be discussed in \S~\ref{results}, the 11 objects marked as `CND'
in Table~\ref{pms_objects_a}, and referred as `candidates' in the text,
could not be observed in our spectroscopic survey.
Hence, their PMS nature remains still uncertain.

\subsection{FLAMES observations}
\label{specfallup}

The two main goals of the spectroscopic observations 
were the detection of the Li{\sc i} $\lambda$6708 
absorption line, a classical youth indicator 
in late-type stars \citep[see, e.g.,][]{Roc02}, and the 
spectral type classification\footnote{The investigation 
on radial velocities and spectroscopic multiplicity 
will be matter of a future paper.}. 
To these aims, multi-object spectroscopy was performed with 
FLAMES\footnote{Fibre Large Array Multi-Element Spectrograph.} 
\citep{Pas02} at the ESO VLT-UT2 
during two observing runs in visiting mode.
The first run was in February 2006, while the second 
one in February 2007.
The weather during both runs was photometric under 
sub-arcsecond seeing conditions. 
We used both the spectrographs fed by FLAMES, i.e. 
GIRAFFE\footnote{Grating Instrument for Radiation Analysis with a Fibre Fed \`Echelle.} 
(in the MEDUSA\footnote{Separate Multi Object.} configuration) and UVES. 
GIRAFFE-MEDUSA allows the simultaneous observation of about 
100 targets with intermediate resolution (about 8,600) in 
the spectral range 6440--7180\AA, while UVES\footnote{UV-Visible Echelle Spectrograph.} 
can access only up to 8 objects at the time and provides 
a resolving power of about 48,000 in the spectral range 4800--6850\AA~. 
Such resolving powers and spectral coverages are ideal to resolve 
the Li{\sc i} line from the nearby Ca{\sc i} $\lambda$6718 line, 
measure Li equivalent widths with an error of less than 
20 m\AA~ on spectra with a S/N ratio higher than 20, and perform a 
reliable spectral type classification.

\begin{figure}
\epsscale{1.0}
\plotone{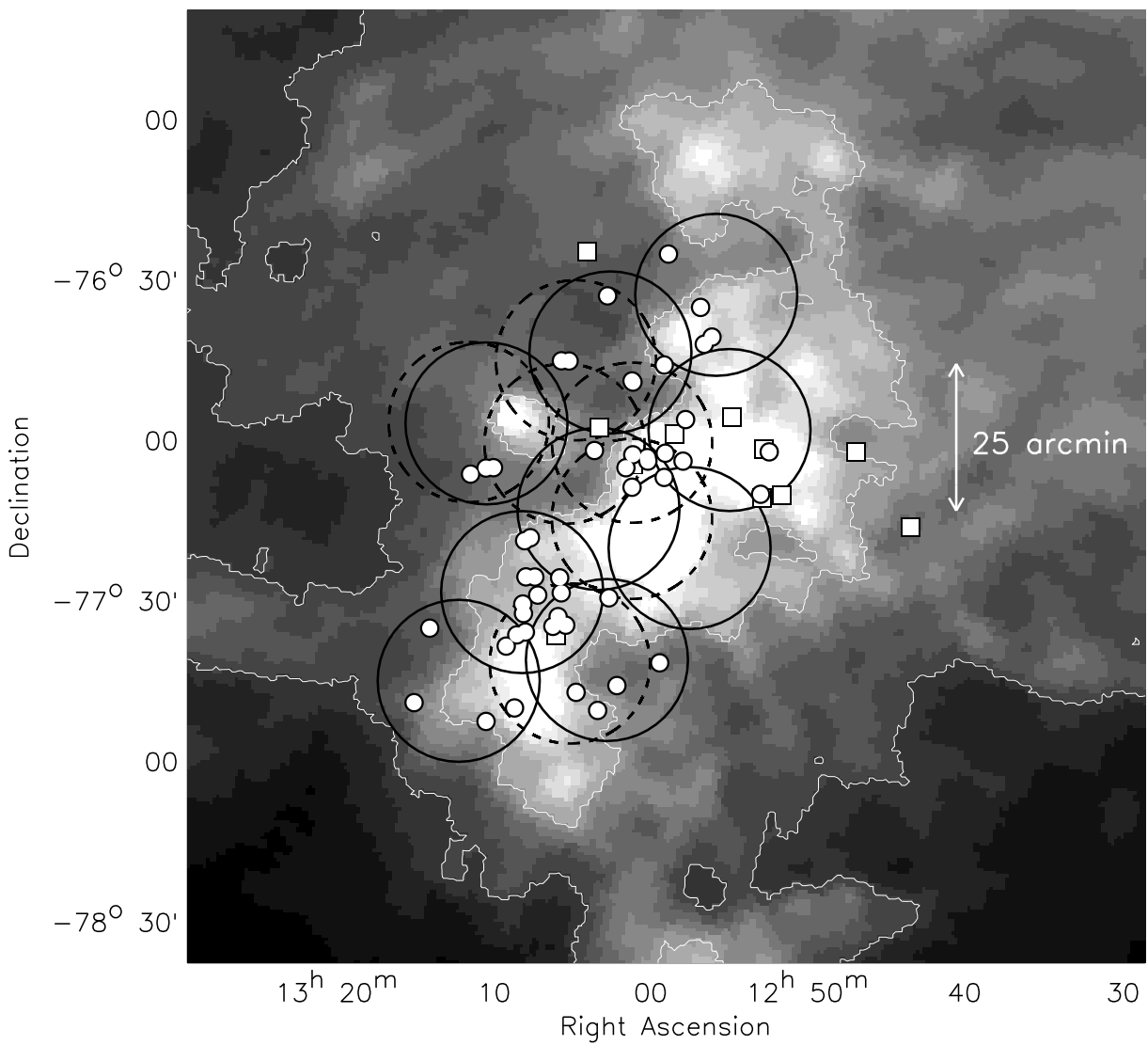}
\caption{Spatial distribution of the confirmed (filled circles) and 
candidate (filled squares) PMS objects in Cha~II 
over-plotted on the map of the IRAS 100$\mu$m dust emission. 
The big circles display the fields covered with FLAMES. 
The fields covered in the first run are represented 
with continuous circles, while those of the 
second run with dashed circles. Each circle is about 
25 arc-min in diameter and North is up and 
East to the left. \label{flames_fields}}
\end{figure}

Figure~\ref{flames_fields} shows the spatial distribution of the FLAMES
fields and Table~\ref{jour_obs} reports the journal of observations.

Despite the relatively large circular field of view of FLAMES, with
approximately 25~arc-min in diameter, it was not possible to assign
a large number of fibers in each configuration to the PMS objects and
candidates because of the rather scattered distribution of the objects
throughout the entire Cha~II region (see Figure~\ref{flames_fields});
on average it was possible to assign only some 5-7 fibers to the main
targets per configuration and in some cases even only 1 or 2. 
Therefore, it was necessary to observe several fields in order to cover
as many PMS objects and candidates as possible. The rest of the fibers
in each configuration were assigned to a few other proposed candidates
(see next section), but most were randomly assigned to field stars.
Although these latter observations are not complete in flux\footnote{The magnitude 
limit of FLAMES in our GIRAFFE-MEDUSA set-up is of the order of R=17.5 and most
of the observed objects are brighter than R=17.},
they allowed to have a check sample and, at the same time, to search
for possible Class~III objects, i.e. objects with no 
prominent infrared excess emission, that might have 
escaped detection in previous surveys. 
In this sense, our multi-object spectroscopic
observations can be considered as an ``unbiased'' Lithium survey.
The number of fibers assigned to the PMS objects and candidates 
and to the field stars are reported in Table~\ref{jour_obs}. 
Note that the magnitude limit of the much higher resolution 
FLAMES/UVES set-up for a S/N$\gtrsim$10 is $R \lesssim$14~mag; 
UVES fibers were then assigned to the brightest PMS 
objects and candidates in our sample.

The typical exposure time per configuration was 3000~sec and, during the
first run, several fields were observed twice or more in order to clean
cosmic rays and increase the S/N of the spectra, which is always 
better than $\sim$20 (see Table~\ref{pms_objects_a}). 

Bias, dark, flat-field and wavelength calibration lamp templates 
were acquired daily under the FLAMES standard calibration 
plan \citep{Mel07}. 
The data reduction, which includes the removal of instrumental 
signatures (i.e. pixel-to-pixel sensitivity variations, 
determination of relative fiber-to-fiber transmission,
fiber localization, fiber spatial PSF modeling, etc) 
and the wavelength calibration, was performed with the on-line 
FLAMES pipeline \citep{Mel07}, whose products are quite 
sufficient for our goals. 
Additional steps for the extraction of the one-dimensional
spectra and sky-spectrum subtraction were performed using 
the {\em apall} package under IRAF. 
The resulting spectra were not calibrated in flux. 
Finally, each one-dimensional spectrum was normalized to the 
continuum by a low-order polynomial fit on the regions of the
spectrum free of absorption lines.
The sampling of the data is 0.2~\AA/pix for GIRAFFE-MEDUSA 
spectra and 0.02~\AA/pix for UVES spectra.

\section{Confirmed PMS objects and candidates}
\label{results}

In total, 41 out the 62 PMS objects and candidates reported in Paper~I were
observed with FLAMES (30 with GIRAFFE-MEDUSA and 11 with UVES), all of which show
the Li{\sc i} line in absorption. Most also display H$\alpha$ in emission.
The 41 objects are flagged in Table~\ref{pms_objects_a} according to their
observation. Fourteen of these objects have been confirmed as Cha~II members
for the first time, while Li{\sc i} has been detected for the first time
in many of the previously known members
\citep[e.g. the H$\alpha$ objects by][and several other IRAS sources]{Har93}.
Among the 21 PMS objects and candidates not observed with FLAMES, 10 were
known to have signatures of genuine PMS objects or they have been spectroscopically
confirmed as members in previous works. We now briefly discuss these
objects:
 i) Three are the Class~I sources IRAS~12500-7658, ISO-CHA\,II\,28, and
    IRAS13036-7644 (BHR~86) discussed in Paper~I. The former is too faint
    for FLAMES spectroscopy and the latter two are not visible in the
    optical;
ii) Another three, namely SSTc2d~J125758.7-770120, SSTc2d~J130540.8-773958, 
    and ISO-CHA\,II\,13 are sub-stellar objects, which are also too
    faint to be observed with FLAMES. These objects were spectroscopically
    confirmed as Cha~II members by \citet{Alc06} and \citet{All07} 
    on the basis of their very late-type spectrum coupled with strong IR 
    excess and signatures of a high level of activity; 
iii) The other four stars are Sz48NE, Sz49, Sz60E, Hn22. 
     The former three have been spectroscopically confirmed 
     as members of Cha~II by \citet{Hug92} using the same criteria 
     as for our observations, i.e. Li{\sc i} line in absorption and late-type 
     spectra, while Hn22 has been detected as a strong H$\alpha$ emitter by \citet{Har93}.
The SED of these 10 objects is typical of PMS 
stars with IR excess emission.

In conclusion, our sample of Cha~II members includes all objects 
for which the Li{\sc i} line has been observed in absorption, either 
in our follow-up spectroscopy or in previous studies, 
and those objects whose young nature is doubtless 
from their strong IR excess or spectral signatures of 
high level of activity. 
We consider as candidates those objects which already passed optical 
or c2d selection criteria based on the inspection of color-color and 
color-magnitude diagrams, but still lack spectroscopy to definitively 
assess their youth.

We then confirm that 51 out of the 62 objects reported
in Paper~I are definitely Cha~II members, while spectroscopy 
is still needed for the 11 remaining objects. 
Note that three of them already passed the c2d selection 
criteria (see Table~7 of Paper~I). The 51 confirmed members
and the 11 candidates are flagged in Table~\ref{pms_objects_a} with ``PMS''
and ``CND'', respectively. Table~\ref{pms_objects_a} reports 
also the H$\alpha$ (EW$_{H\alpha}$) and Li{\sc i} (EW$_{LiI}$) 
equivalent widths of the confirmed PMS objects, 
which were measured in the background-subtracted and 
continuum-normalized spectra using a Gaussian fitting of 
the line profile. The continuum was defined 
by fitting a low order spline to a pre-selected spectral region 
between the H$\alpha$ and Li{\sc i} lines. 
For this purpose we used the $\sim$20\AA~ continuum window between 
6680 and 6700\AA, which is not affected by strong lines in K/M type objects. 
Our EW$_{LiI}$ measurements agree within 0.15\AA~ with those reported 
by \citet{Hug92} for the previously known Cha~II members. Note that 
such RMS difference is mainly dominated by the EW$_{LiI}$ errors
resulting from the \citet{Hug92} low-resolution spectra, which are 
on the order of 0.1-0.15\AA.

Note also that the EW$_{LiI}$ values of the confirmed PMS objects
are typical of low-mass PMS stars in other star-forming regions 
\citep[0.3-0.8\AA;][and references therein]{Leo07}. Though the strength 
of the Li{\sc i} line and the relative uncertainty are expected to vary 
with spectral type, the EW$_{LiI}$ values in our sample are stronger than 
those measured for Pleiades stars of the same spectral type. This criterion
is widely used to single out bona-fide low-mass PMS stars 
\citep{Cov97a, Mar99b, Wic00, Alc00}.

\begin{figure}
\includegraphics[angle=0,scale=0.75]{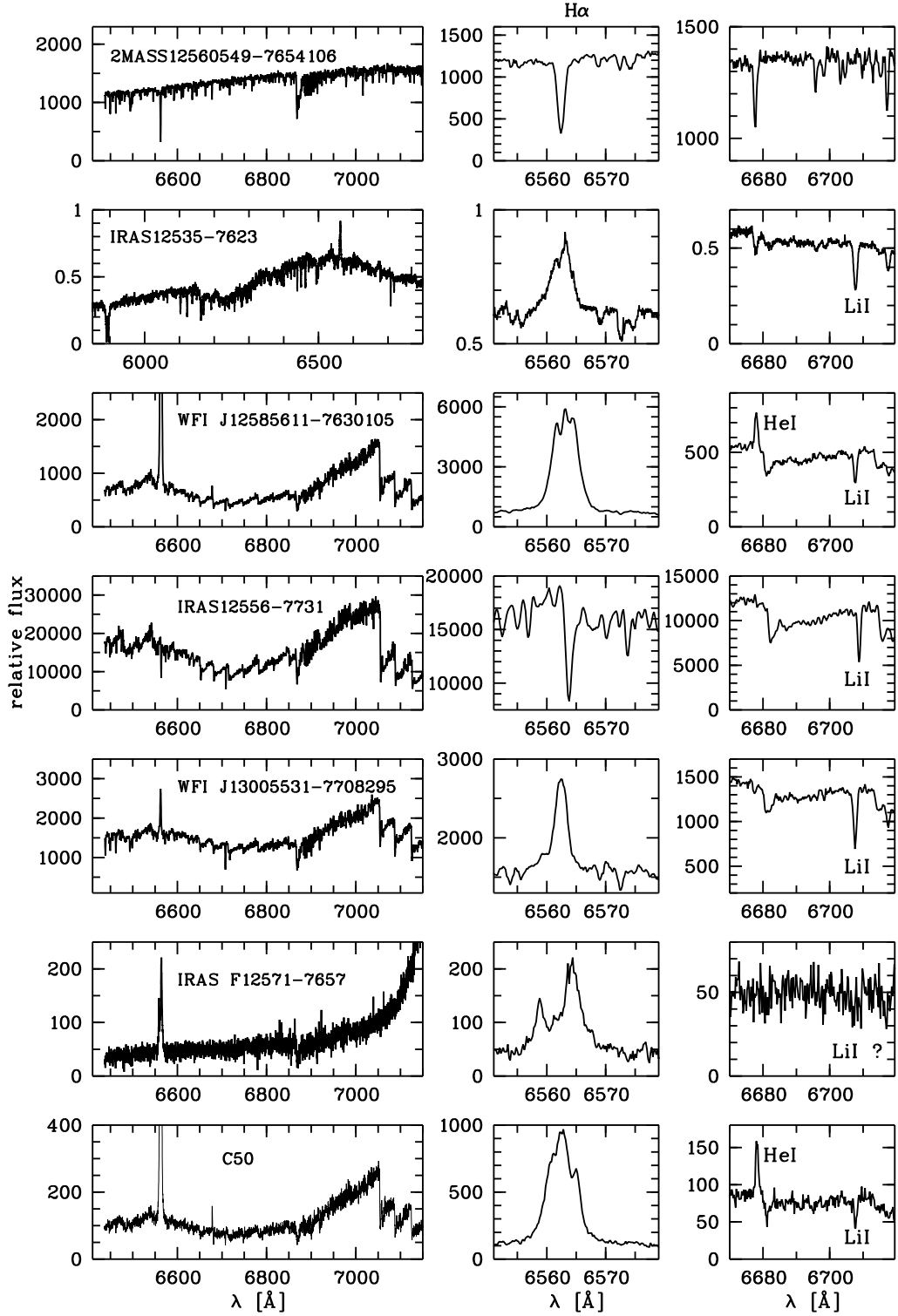}
\caption{
Optical spectra of objects selected as PMS star candidates
in Paper I. The left panels show the complete spectral range. 
For the GIRAFFE-MEDUSA spectra the wavelength range is from 
6450~\AA~ to 7150~\AA, while for the two UVES spectra, 
corresponding to IRAS~12535-7623 and IRAS~F13052-7653N, the 
wavelength coverage shown in the figure is from 5850~\AA~ to 
6800~\AA. The central and right panels show the H$\alpha$ and 
Li{\sc i} spectral regions, respectively.
\label{spectra} }
\end{figure}

In Figure~\ref{spectra} examples of FLAMES spectra of 20 objects selected
as PMS candidates are shown; among them we show the 14 members confirmed
for the first time. The strong Li{\sc i} $\lambda\lambda$6708~\AA~ absorption
is clearly visible, with most of them displaying H$\alpha$ in emission.
Among the 20 spectra, we also show six lacking the Li{\sc i} line; these
are 2MASS~12560549-7654106, 2MASS~13102531-7729085, WFI~J12591881-7704419,
WFI~J12592348-7726589, WFI~J13014752-7631023, and WFI~J13071960-7655476.
The former two were proposed as member candidates by \citet{You05} but
both were rejected on the basis of the optical 
criteria by \citet{Spe07}; note that these two 
objects show H$\alpha$ in absorption. The latter four
were selected by \citet{Spe07} with two of them displaying H$\alpha$
emission, all this in agreement with the H$\alpha$ photometric index
for these objects reported by \citet{Spe07}. Because of the definitive
lack of Li{\sc i} absorption, these six sources are not PMS objects and
are thus unrelated to Cha~II. Note that about 50\% of the candidates
reported by \citet{Spe07} in their Table~6 are spectroscopically confirmed
to be Cha~II members, in line with their conclusions.

Several of the spectroscopically confirmed PMS objects show the
He~I $\lambda\lambda$6678~\AA~ line in emission. This line is the singlet
counterpart of the He~I $\lambda\lambda$5876~\AA~ and indicates strong
chromospheric activity, consistent with the young age of these objects.
Moreover, the different H$\alpha$ emission line profiles, typical of
PMS stars, can be appreciated.
For instance, both IRAS~F12571-7657 and SSTc2d~J130521.7-773810  show
strong H$\alpha$ emission and may well be cases of veiled objects.
Moreover, the spectrum of SSTc2d~J130521.7-773810 exhibits the [NII]
and [SII] forbidden emission lines at $\lambda\lambda$6584 and  
$\lambda\lambda$6731. As shown in Paper I (Figure~7),
these two objects show substantial IR excess in their SEDs.  

The presence of strong Li{\sc i} absorption and the weak H$\alpha$ 
emission in the spectrum of ISO-CHA\,II\,29  allow us to classify 
this object as a weak-line T Tauri star. This, together with its 
peculiar spectral energy distribution, which rises beyond 
24~$\mu$m (see Paper~I), confirms ISO-CHA\,II\,29 as a ``transition'' 
object.

Several other objects, proposed as possible members in previous works
\citep{Vuo01, Per03, You05, Spe07}, were also observed, but not confirmed 
as members (see Table~\ref{tab:rej}). For instance, ISO-CHA\,II\,73, 
ISO-CHA\,II\,110, ISO-CHA\,II\,32, ISO-CHA\,II\,91, already rejected by 
the \citet{Spe07} optical criteria, have neither Li{\sc i} nor H$\alpha$ 
emission in their spectra. 
The spectra of these objects are typical of K-M giants unrelated 
to Cha~II (cf. Figure~\ref{rejected}). For our purposes, this type of 
objects are then counted as field stars in Table~\ref{jour_obs}.

\begin{figure}
\includegraphics[angle=0,scale=0.75]{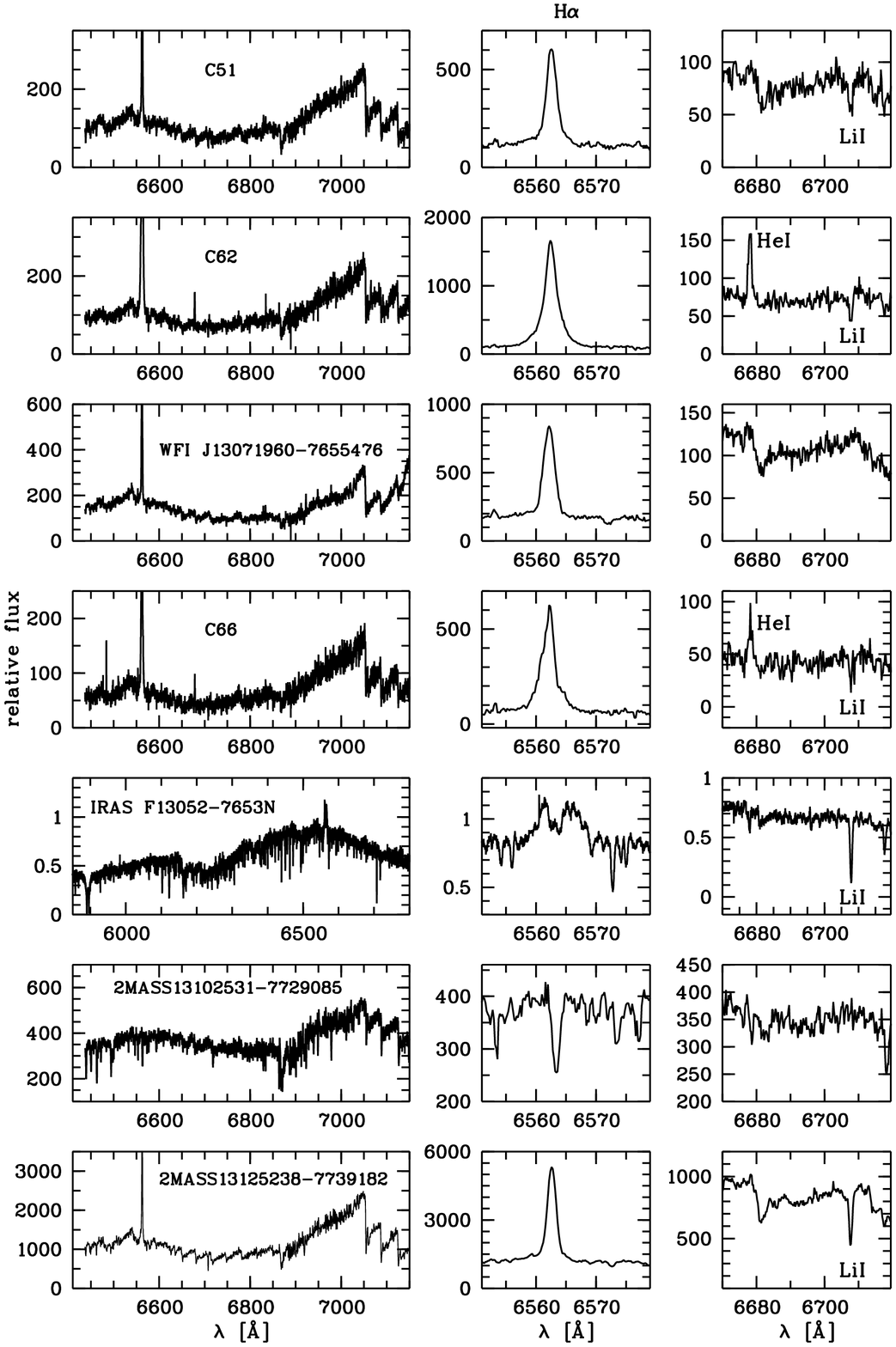}
~~\\
Fig.~\ref{spectra} - Continued
\end{figure}

\begin{figure}
\includegraphics[angle=0,scale=0.75]{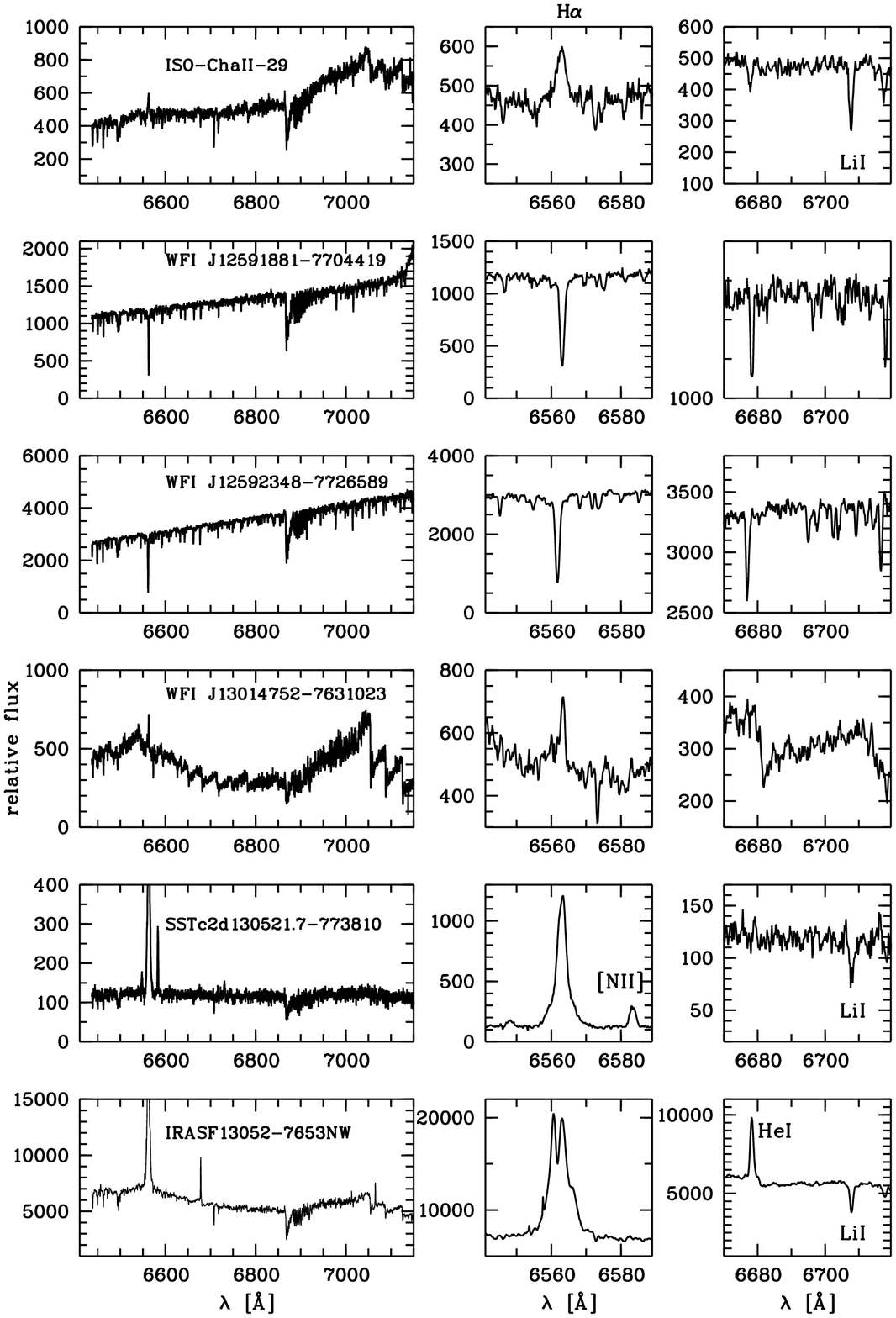}
~~\\
Fig.~\ref{spectra} - Continued
\end{figure}


\begin{figure}
\includegraphics[angle=0,scale=0.78]{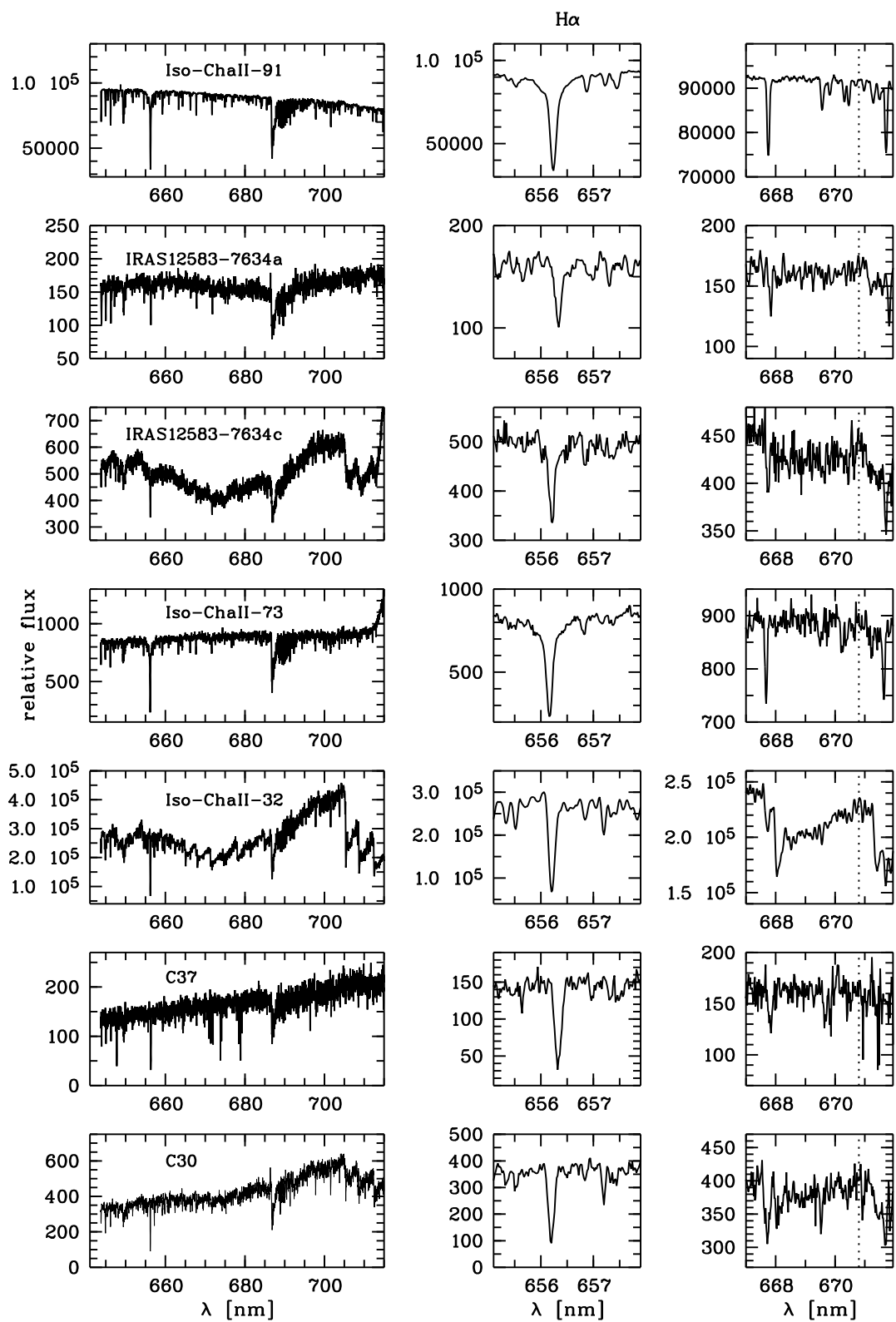} 
\caption{Examples of optical spectra of sources selected as
Cha~II candidates members in previous surveys, but rejected 
by our spectroscopic criteria. 
The dotted line in the rightmost panels marks the position
where the Li{\sc i} absorption line should be. Note that
none of these objects shows the line. \label{rejected}}
\end{figure}

\begin{figure}[!h]
\includegraphics[angle=-90,scale=0.62]{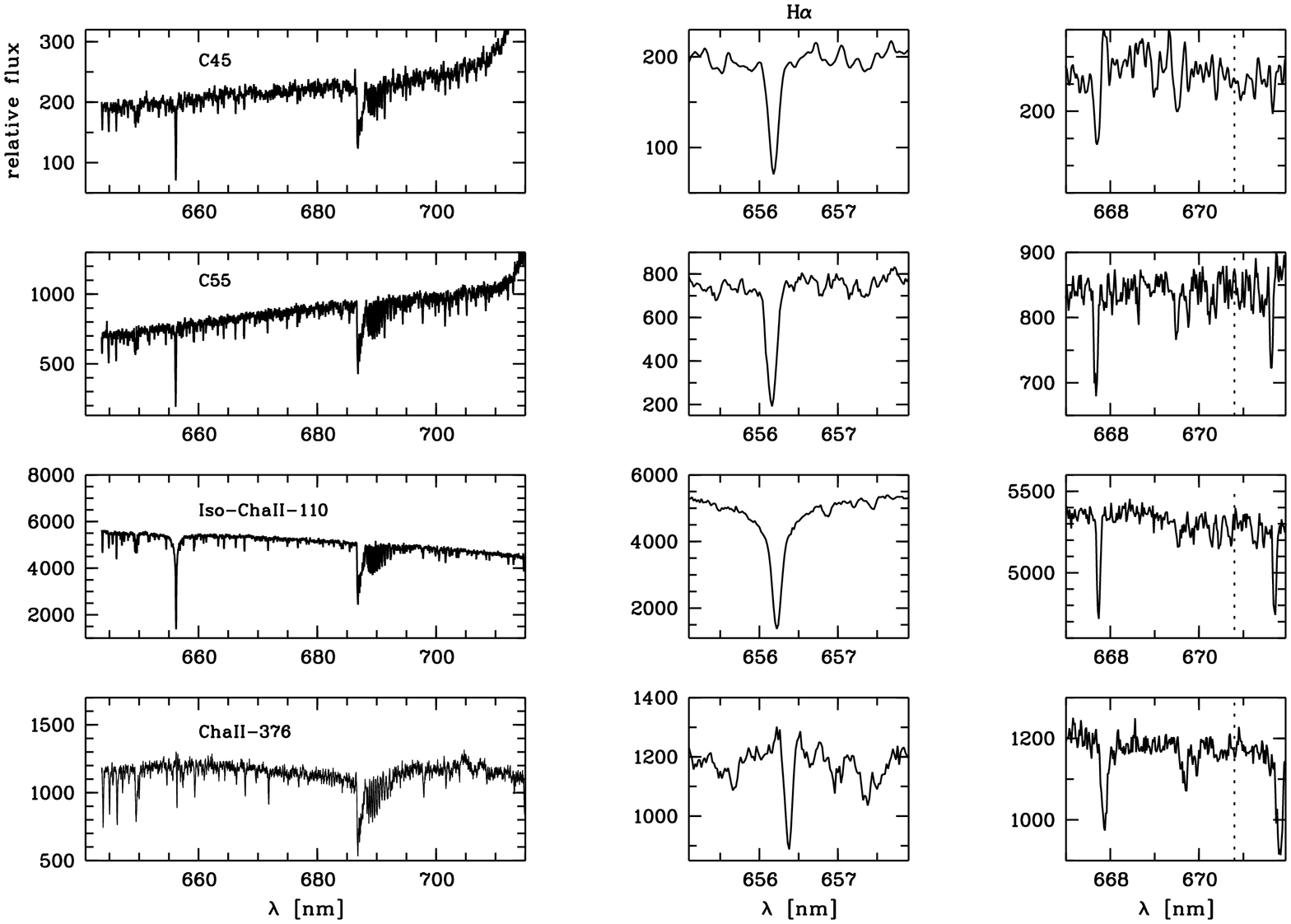}
~~\\
Fig.~\ref{rejected} - Continued
\end{figure}


The fraction of Class~II to Class~III objects in ChaII is the largest 
in the whole c2d cloud sample. Unlike other T associations, the PMS 
population in Cha~II appears dominated by Class~II sources ($\sim$60\%), 
which outnumber Class~III sources by a factor $\sim$2 (see Table~1 of Paper~I). 

An important result of our FLAMES observations is that no new Class~III 
PMS objects were found among the $\sim$2000 field stars observed. All these 
stars lack detectable Li{\sc i} or H$_\alpha$ emission in their spectra. 
By using the catalog of optical sources in the Cha~II direction by \citet{Spe07}, 
we estimate that these $\sim$2000 stars represent $\sim$30\% of the total 
field stars with $R<$17.5 in the area covered by our FLAMES observations. 
Therefore, unless all the Class~III sources are located behind the cloud, 
which is very unlikely, this result supports the completeness of the sample
 of PMS objects and candidates reported in Paper~I.

\section{Stellar parameters}
\label{par}

With the sample of certified PMS objects and candidates, we can proceed 
to their characterization. In this section we outline the procedures
used to determine the physical parameters.

\subsection{Spectral types}
\label{SpecTyp}

A spectral type was assigned to each observed PMS star in Cha~II 
by using a variation of the ROTFIT code developed by \citet{Fra03} 
under the IDL environment. The general idea of the method is to 
recover the spectral type of an object by comparing its spectrum 
with a grid of templates with similar resolution. To this aim, the 
intermediate-resolution standard templates provided by \citet{Mar99}, 
\citet{Haw02}, \citet{LeB03}, \citet{Val04}, and \citet{Boc07}, as 
well as  the high-resolution standard libraries from \citet{Cin04} 
and \citet{Bag03} were used to perform the spectral type 
classification of the GIRAFFE-MEDUSA and UVES spectra, respectively.

As demonstrated by many authors \citep{Luh99, Luh03, Gui06, Gan07},
because of the presence of TiO and VO absorption bands which
are sensitive to the surface gravity, most of the M-type PMS
objects show spectroscopic features that can be better reproduced
by averaging dwarf and giant spectra with the same spectral type.
Thus, in performing the spectral type classification of the young 
objects in Cha~II, both dwarf and giant templates were included in our 
grids of reference spectra. Furthermore, for spectral types later than 
M0, we also included \emph{ad hoc} templates obtained by averaging the 
collected spectra of dwarfs and giants with the same spectral type.
 
The template spectra were first normalized to the continuum 
through a low-order polynomial fit, in the same way as the target 
spectra (cf. \S~\ref{specfallup}). 
The standard spectra were then shifted in wavelength to the radial 
velocity of the target. Finally, a chi-square minimization procedure 
was applied to choose the template that best fits the observed 
spectrum.

From an inspection of the residual we estimated an accuracy on the 
resulting spectral types within one spectral sub-class for objects earlier 
than $\sim$K7 and about half spectral sub-class for later objects.

We could perform a reasonable spectroscopic classification for
all the sources observed with FLAMES, except for Sz~47, IRAS~F12571-7657 
and SSTc2d~J130521.7-773810. The former object was classified as M0 
by \citet{Hug92}, but  we observe signatures of a much earlier spectral 
type; for instance, a strong He~I $\lambda\lambda$6678~\AA~ line in 
absorption and a rising continuum with shorter wavelength. Already in 
the low-resolution spectrum presented by \citet{Hug92} the He~I absorption 
is clearly visible. Such  spectral features resemble those of an early 
type object. This effect might be due to the strong interaction
of the star with the disk. In fact Sz~47 is another case of veiled
star in Cha~II. The spectra of IRAS~F12571-7657 and SSTc2d~J130521.7-773810
are rather noisy and although we manage to perform a fit of a K3
and K6 templates respectively, from the almost pure-continuum 
appearance of their spectra, they may be classified as continuum-type 
or ``C-typ'' objects.

The spectral types are provided in Table~\ref{pms_objects_a}. \citet{Hug92}
reported spectral types for 20 stars in Cha~II; 17 of these were
observed by us with FLAMES. In most cases, we find a good agreement
of the spectral types reported by these authors within one spectral
sub-class. In a few cases (e.g. Sz~47 and Sz~62) larger residuals,
up to two spectral sub-classes, are found. For these objects the
spectral type determination becomes more uncertain because of veiling.

Among the sample of spectroscopically confirmed PMS objects there
are some with spectral types M5 or later, which correspond
to objects very close or below the Hydrogen burning limit.
These include objects confirmed in this work and two very low-mass
BDs discovered by \citet{All06b}. For the latter we adopt the spectral 
types and temperatures recently determined by \citet{All07}. Further 
discussion of these objects is presented in \S~\ref{substellar_objs}.

\subsection{Effective temperatures}
\label{Teff}

For the confirmed PMS objects, the effective temperature (T$_{\rm eff}$)
was determined by using the tabulations of temperature as a function of
spectral type proposed by \citet{Ken95} and \citet{Luh03}.
Consistently with our spectral classification, the dwarf temperature
scale by \citet{Ken95} was considered for objects earlier than M0,
while for later objects we adopted the tabulation by \citet{Luh03}, which
is intermediate between dwarf and giant scales.
The errors on T$_{\rm eff}$ were estimated considering a mean uncertainty
of one sub-class on the spectral type classification.

A different approach was followed for the objects lacking spectroscopy.
In these cases, the effective temperature and extinction were 
simultaneously derived by fitting a grid of reference SEDs to their SED, 
as prescribed by \citet{Spe07}, but using the \citet{Wei01} extinction 
law (see also \S~\ref{ext}).
To this aim, we used the observed fluxes presented in Paper~I and 
restricted the fit to the short-wavelength portion of the SED 
($\lambda \leq \lambda_{J}$), which is less contaminated by possible 
IR excess.
When possible, these temperature values were also checked by
using the T$_{\rm eff}$  vs. ($m_{856}-m_{914}$) calibration relation
determined by \citet{Spe07}; the ESO-WFI ($m_{856}-m_{914}$) color 
index is sensitive to the effective temperature for very cool objects 
(2000~K$\lesssim$T$_{\rm eff}\lesssim$3800~K) because 
the medium-band filter centred at 856\,nm covers important TiO 
absorption features that deepen with decreasing temperature, 
while the one centred at 914 nm lies in a 
wavelength range relatively featureless 
in late-type objects. In the case of high interstellar extinction 
(A$_V \gtrsim$5~mag) the temperature derived from this relation 
may be underestimated because the ($m_{856}-m_{914}$) 
index becomes larger than in the absence of extinction. 
The two methods yield consistent results and turn out to
be accurate within 200\,K relative to the spectroscopic temperature
estimated for the objects observed with FLAMES. 

The effective temperatures for PMS objects and candidates in Cha~II
are reported in Table~\ref{pms_objects_b}.

\subsection{Extinction}
\label{ext}

The extinction has been determined for each individual PMS object
using the relation $A_V = 4.61 \times E(R-I)$ obtained from the extinction
law by \citet{Wei01} for R$_V$=5.5 (hereafter WD5.5); indeed, investigations
on the properties of the interstellar medium in the Chamaeleon clouds
have demonstrated that the R$_V$ ratio is unusually high (i.e. $R_V$=5-6)
in the densest parts of the clouds \citep{Cov97, Luh06}. For details on
this issue we refer the reader to Paper~I.
The $E(R-I)$ color excess was derived from the compilation of colors as
a function of spectral type proposed by \citet{Ken95} and \citet{Luh03}
for spectral types earlier than M1 and later than M0, respectively.
The spectral types determined as explained in \S~\ref{SpecTyp} and
reported in Table~\ref{pms_objects_a} were used to this aim.
For objects lacking the $R$ or $I$ band magnitudes, also because
of saturation, we used the intrinsic $(J-H)$ color versus spectral
type relation by \citet{Ken95}, and determine the extinction as
$A_V = E(J-H) / 0.10$, also using the WD5.5 extinction law. This was
the case for some IRAS sources which are very bright also in the
optical (see \S~\ref{sec_hrd}). This latter approach was also used for
C~41 and Sz~49, for which the $(R-I)$ color is strongly affected by the
intense H$\alpha$ emission and veiling. The uncertainties on $A_V$, 
derived from the uncertainties on spectral classification, range 
from 0.1 to less than 1~mag.

For the objects lacking spectroscopy, the $A_V$ and T$_{\rm eff}$ 
values were derived simultaneously following the SED minimization 
procedure prescribed by \citet{Spe07}.
We used the observed fluxes presented in Paper~I and also adopted
the WD5.5 extinction law. As shown in \citet{Spe07}, this method 
provides $A_V$ values consistent with those derived from spectroscopy
within 1.5\,mag. Such uncertainty translates into an average uncertainty 
of 0.15~dex in the logarithm of luminosity for the PMS candidates 
in Cha~II.

The extinction at each wavelength, $A_{\lambda}$, has been then determined 
both for the PMS objects and the candidates also using the WD5.5 extinction 
law. As discussed in Paper~I, our extinction determinations are in general 
agreement with the values reported by the large-scale extinction maps of 
the Cha~II cloud. 

\subsection{Stellar luminosities}
\label{lum}

Because of the presence of circumstellar material, the SEDs of
young objects may be affected by the presence of IR excess and hot
continuum emission from the boundary layer between the disk and
the central star, whose intensities depend on the evolutionary
stage of the star. In order to separate the properties of the
central star alone and the disk, their contributions to the
``total'' luminosity of the system must be determined.

The ``total'' luminosities of the young objects in Cha~II,
obtained by direct integration of the dereddened SEDs, have been
already reported in Paper~I. In that paper we concentrate on the
disk contribution, which provides an interesting diagnostic for
the study of the accretion activity of PMS objects.
Here we focus on the contribution of the central star alone,
which allows us to place the Cha~II members and candidates
on the HR diagram and hence derive their masses and ages.
The stellar luminosities and radii for the Cha~II population were
computed following the procedure described in \citet{Spe07}.
We compared the dereddened SED of each object with a reference SED
of the same temperature scaled to the \objectname{Cha~II} distance
\citep[d=178~pc,][]{Whi97}. The grid of reference SEDs was built
using the model spectra by \citet{Hau99} and \citet{All00} as explained
in \citet{Spe07}.
By minimizing the flux differences between the dereddened SED and
the reference one, we determine the stellar radius (R$_\star$) and hence,
the photospheric luminosity (L$_\star$). The minimization procedure was 
performed, as above, on the short-wavelength portion of the SED 
($\lambda \leq \lambda_{J}$), in order to avoid contamination from 
possible IR excesses. 
Errors on luminosities have been estimated taking the uncertainty 
on spectral classification and extinction (i.e. $\sim$1 and 2 spectral 
sub-class for objects with and without spectroscopy, respectively), as well
as on the distance to Cha~II \citep[$\sim$20~pc;][]{Whi97} into account. 
We also checked that using extinction laws other than the WD5.5, 
such as those by \citet{Sav79} and \citet{Car89}, does not affect 
significantly our luminosity estimates. The  R$_\star$ and L$_\star$ 
values and relative uncertainties are reported in Table~\ref{pms_objects_b}.

\subsection{HR diagram}
\label{sec_hrd}

\begin{figure}
\epsscale{1.0}
\plotone{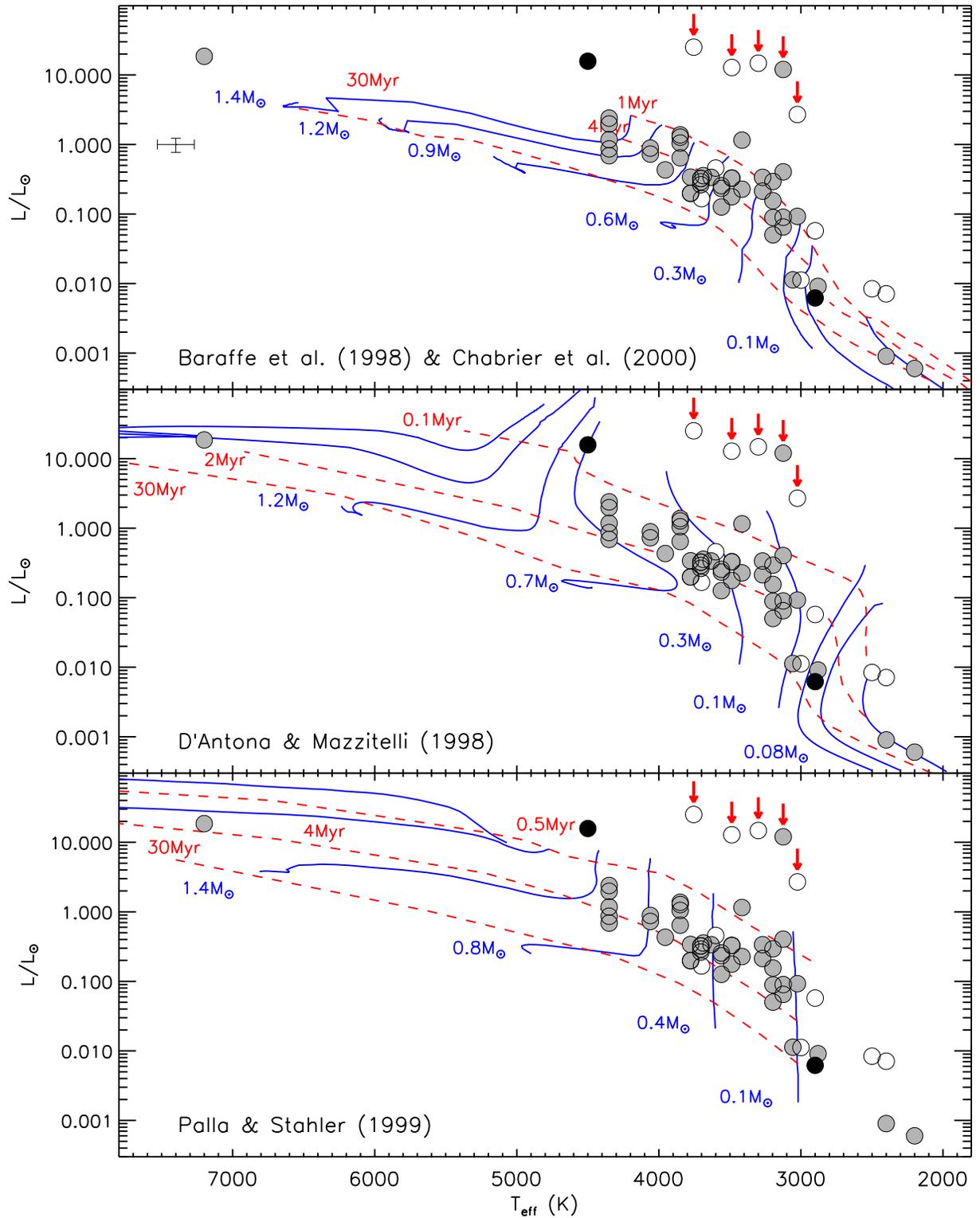}
\caption{Luminosity versus temperature diagram for the PMS objects
        (gray filled circles) and candidates (open circles)
     over-plotted on three PMS evolutionary models as
     indicated in each panel. The black filled circles represent
     the objects classified as Class~I sources. The arrows indicate
     the over-luminous sources (see \S~\ref{sec_hrd}). 
     Mean errors on luminosity and temperature are shown 
     in the upper left. \label{hrd}}
\end{figure}

The positions on the HR diagram of the PMS objects and candidates in Cha~II
are presented in Figure~\ref{hrd} over-plotted on three sets of evolutionary
tracks. Except for a few cases discussed next, the vast majority of
the confirmed PMS objects and candidates fall in regions of the HR
diagram consistent with very young objects (age$\approx$4~Myr), down
to the BD regime. The luminosity and temperature of five objects are,
however, inconsistent with those of PMS stars in Cha~II; these objects
are marked with arrows in Figure~\ref{hrd} and correspond to the five IRAS
sources discussed in \S~5.4 of Paper~I.
These sources are IRAS12416-7703, IRAS12448-7650,
IRASF12488-7658, IRAS12556-7731 and IRAS12589-7646, which would be
over-luminous in the HR diagram. Note that one of these, i.e. IRAS~12556-7731,
has been spectroscopically confirmed to be a PMS object (\S~\ref{results}).
Possible reasons for the inconsistent position of these objects
on the HR diagram may be an erroneous estimate of temperature and 
consequently of extinction. It is worth to mention, however, that all 
these IRAS sources are very bright ($13<R<14$) and are saturated in 
the $R$ and $I$ band WFI images \citep[c.f.][]{Spe07}. As explained in 
\S~\ref{ext}, the extinction in these cases was determined on the basis 
of the $(J-H)$ color index. Note that none of the five IRAS sources 
show very strong IR excess, with most of them expected to be Class~III
sources. Therefore, the methods applied here to determine their temperature
and extinction are reliable. In addition, the visual extinction for
these objects is less than about 4~mag. Note that at least in
the case of IRAS12556-7731, the spectral type and temperature are fairly
well constrained from its spectrum. Yet, this object is over-luminous at 
the same degree as the other four IRAS sources.

Interestingly, while being scattered in the region (see Figure~10 in Paper~I),
the five IRAS sources seem to follow the form of an isochrone on the HR diagram.
Thus, another possible explanation for the inconsistency may be that the
distance for these five IRAS sources has been overestimated. In order for
them to have a  position on the diagram consistent with PMS objects 
(age$\approx$4~Myr), their distance should be on the order of 30~pc.  
Alternatively, they could be foreground dwarfs, unrelated
to Cha~II and the apparent IR excess by which some of them were
selected might simply be due to cool companions. More spectroscopic 
data are necessary in order to shed light on the nature of these 
outliers. The immediate consequence is that their mass and age 
cannot be determined here.

\subsection{Masses and ages}
\label{sec_mass_age}

We have derived masses and ages of the Cha~II members and candidates
by comparison of the location of the objects on the HR diagram
with the PMS tracks by \citet{Bar98} \& \citet{Cha00},
\citet{DAn98} and \citet{Pal99}. Since stellar evolutionary models are rather
uncertain, particularly in the low-mass and sub-stellar regimes, the use
of different evolutionary models allows us to estimate the uncertainties 
of the stellar parameters due to modeling alone. In Table~\ref{tab:par2} 
we report the masses and ages for all the confirmed Cha~II members and 
candidates whose position on the HR diagram is consistent with membership 
to the cloud.

\begin{figure}[!h]
\epsscale{1.0}
\plotone{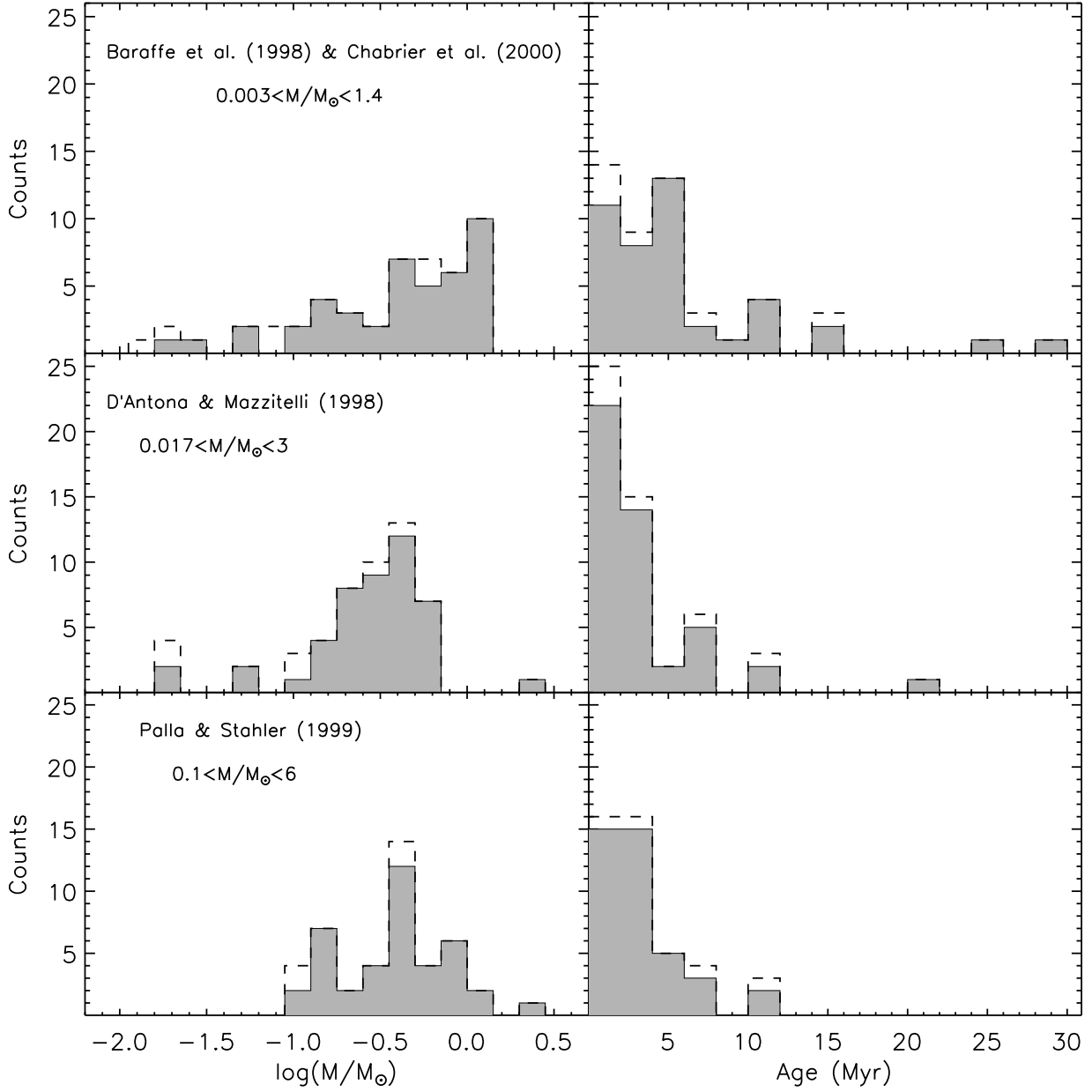}
\caption{Histograms of mass (left panels) and age (right panels) of the PMS
      objects and candidates in Cha~II resulting from different evolutionary
      tracks and isochrones as indicated. The gray histograms display
      the distribution for the certified PMS objects only,
      while the dashed histograms are obtained when adding 
      the candidates. The mass range covered by each set of 
      evolutionary models is indicated in the left panels. 
      \label{mass_age}}
\end{figure}

Most of the Cha~II PMS members have masses in the range
$0.015 \leq M \leq 1 M_{\odot}$; IRAS~12496-7650 (DK~Cha) is the only
intermediate mass star in the cloud.  The mass distribution of the
Cha~II members, which is shown in Figure~\ref{mass_age} (left panel),
peaks between 0.3 and 0.4~$M_{\odot}$ and the mean mass ranges from
0.4 to about 0.6~$M_{\odot}$, depending on the adopted evolutionary
track (c.f. Table~\ref{tab:SF}).
The resulting average mass is 0.5$\pm$0.1~$M_{\odot}$, with the
error representing the uncertainty due to modeling. It is important
to note that these values are basically the same when including the
candidates. Note also that the resulting values for the YSOs alone,
also reported in Table~\ref{tab:SF}, fall in the same range.
The average mass is similar to the values obtained for
Cha~I ($\sim$0.45 $M_{\odot}$) and Taurus (0.50 $M_{\odot}$)
populations\footnote{In order to compare these values,
we have consistently derived the average mass of the young
populations using data by \citet{Luh04} for Cha~I and \citet{Bri02}
for Taurus, respectively.}, indicating that a value of the order 
of 0.5$M_{\odot}$ might be a characteristic average mass for objects 
in T associations.
 
 
The age distributions, resulting from the adopted models, are presented
in the right panels of Figure~\ref{mass_age}. The vast majority of PMS objects 
and candidates are younger than 4~Myr. The mean age of the Cha~II population 
ranges from 3 to 6~Myr depending on the adopted model (Table~\ref{tab:SF}).
Independently of whether or not the candidates are included, the average
age is 4$\pm$2~Myr with the error coming from the modeling alone.
Within the uncertainties, this value is in agreement with that derived
by \citet{Cie05}.
The mean age for the YSOs alone, also reported in Table~\ref{tab:SF},
is lower ($\sim$2~Myr), though not significantly; this is expected since 
objects with strong IR excess are representative of an earlier 
evolutionary phase.

The majority of the objects have ages of 3-4 Myr, depending on the 
models. Therefore, star formation proceeded in a short time, producing 
a coeval population. If the star formation in Cha~II occurred very recently, 
say within the last Myrs, the circumstellar disks of the PMS objects 
in this region may have not yet had the time to evolve into optically 
thin disks. In other words, this may indicate that star formation in 
Chamaeleon has occurred rapidly a few million years ago.

\subsection{Stellar parameters for Class~I sources}
\label{classI}

The analysis of the IR properties of the young population in Cha~II presented 
in Paper~I shows that only three Class~I sources exist in the cloud, namely 
IRAS~12500-7658, IRAS~12553-7651 (or ISO-CHA\,II\,28) and IRAS~13036-7644 
(or BHR~86).

In these cases we can not apply the above procedures for the determination
of the stellar parameters. Indeed, the large extinction coupled with strong
IR excess due to the thick envelope/disk, makes it difficult to disentangle
the properties of the central object from those of the circumstellar regions.
Since most of the emission of Class~I sources is in the IR, one way to
determine their properties is by fitting a disk accretion model to their SED.
From such fit it is possible to provide an estimate of the temperature
and other physical parameters of the central object. This has been done for
the Cha~II Class~I sources in Paper~I by using the SED models for young stellar
objects by \citet{Rob06}.
Thus, for IRAS~12500-7658 and ISO-CHA\,II\,28 we report in Table~\ref{pms_objects_b}
the values obtained in Paper~I, which have been also used to place them on
the HR diagram (c.f. Figure~\ref{hrd}). For IRAS~13036-7644 the SED modeling
results are unreliable because it is not detected neither in the optical nor
in the near-IR; this object has been matter of focused c2d observations which
will be presented in a future paper.

We warn the reader that ISO-CHA\,II\,28 is not detected in the optical and
only a flux upper limit is available in the $J$-band from the 2MASS catalog.
Thus, visual extinction and other stellar parameters resulting from the SED
modeling must be taken with care.

The stellar parameters estimated for IRAS~12500-7658 point toward a very
cool low-mass object. Since this object has been detected in the optical
survey by \citet{Spe07}, we could independently estimate its stellar parameters
using the techniques described in \citet{Spe07}. 
The WFI ($m_{856}-m_{914}$) index of IRAS~12500-7658 indicates an effective 
temperature of 3000$\pm$200~K.
Assuming this temperature, we estimate a visual extinction  A$_V=3.93$,
an object luminosity of 0.0067~L$_{\odot}$, an integrated luminosity of
0.9~L$_{\odot}$ and a stellar mass of about 0.06~M$_{\odot}$. 
These values are roughly consistent with those derived by using the 
\citet{Rob06} disk models. However, we remind that the effective 
temperature estimates from the WFI ($m_{856}-m_{914}$) index may be 
significantly affected by extinction, therefore they are rather uncertain.

\section{Objects near and below the sub-stellar limit}
\label{substellar_objs}

There are several objects which fall close to or below the Hydrogen
burning limit. Three, namely SSTc2d~J125758.7-770120, ISO-CHA\,II\,13,
and SSTc2d~J130540.8-773958, are spectroscopically confirmed to be
sub-stellar \citep{Jay06, Alc06, All07}\footnote{Note that
for ISO-CHA\,II\,13 \citet{All07} derived a slightly higher temperature
than used here. Since the average temperature derived from the values
reported by \citet{Alc06} and \citet{All07} still places the object
in the BD domain, we consider it as sub-stellar.},
while four (WFI~J12533662-7706393,
WFI~J12583675-7704065, WFI~J13005297-7709478 and WFI~J13031615-7629381)
are candidates to be sub-stellar.

On the other hand, five sources (C41, C50, C62, C66 and WFIJ12585611-7630105)
are spectroscopically confirmed to be very low-mass PMS stars close to the
Hydrogen burning limit, while the temperature estimate of IRAS12500-7658
is consistent with an object close to this limit (\S~\ref{classI}). 

Considering only the certified PMS objects, the fraction of sub-stellar 
objects relative to PMS stars would be on the order of 6\%, i.e. lower than
in other T associations. 
It is clear that this result may change slightly because of the fact that 
several of the  objects fall just above or below the Hydrogen burning limit 
and because of the uncertainties in spectral types and of the difference in 
the adopted PMS evolutionary models. In whatever case, the fraction may be 
slightly lower than, but very close to the values reported for other 
T associations \citep[12-14\%,][]{Bri02,Lop04} and lower than derived
for OB associations \citep[$\sim$26\%,][and references therein]{Hil00, Bri02, Mue02}.
This points toward a sub-stellar IMF for Cha~II similar to that found in
other T associations. More discussion on the Cha~II IMF is deferred
to \S~\ref{imf}.

\section{The star formation in the Chamaeleon II dark cloud}
\label{SF}

Fundamental physical parameters, i.e. mass and age of the individual PMS objects,
have been now determined; moreover, as discussed in Paper~I, the census of Cha~II
members is rather complete down to the sub-stellar
regime \citep[$M \approx 0.03~M_{\odot}$,][]{Alc07}.
Then, the implications on the star formation in the cloud can be investigated on
the basis of usual quantities such as the star formation efficiency, the trend of
the mass spectrum and the star formation rate. Some aspects of the star formation
history of the cloud are also discussed.

\subsection{A first guess on the IMF in Cha~II}
\label{imf}

Despite the considerable theoretical and observational work establishing
the general form of the IMF \citep{Sca86, Sca98, Kro01, Kro02, Rei02, Cha03},
the question whether there are differences in the IMF under different
star-forming conditions is still an open issue. This is particularly
controversial in the very-low mass and sub-stellar regimes.
Recent studies \citep{Hil00, Luh00, Bri02, Pre03, Mue03} show that the
IMF remains approximately flat or shows a moderate decline in the BD
regime. Nevertheless, there is no systematic indication of such a
turnover. Other regions have been found where the IMF appears to rise
below the sub-stellar limit \citep{Hil00, Luh00, Bri02, Pre03, Mue03}.

\begin{figure}[!h]
\epsscale{1.0}
\plotone{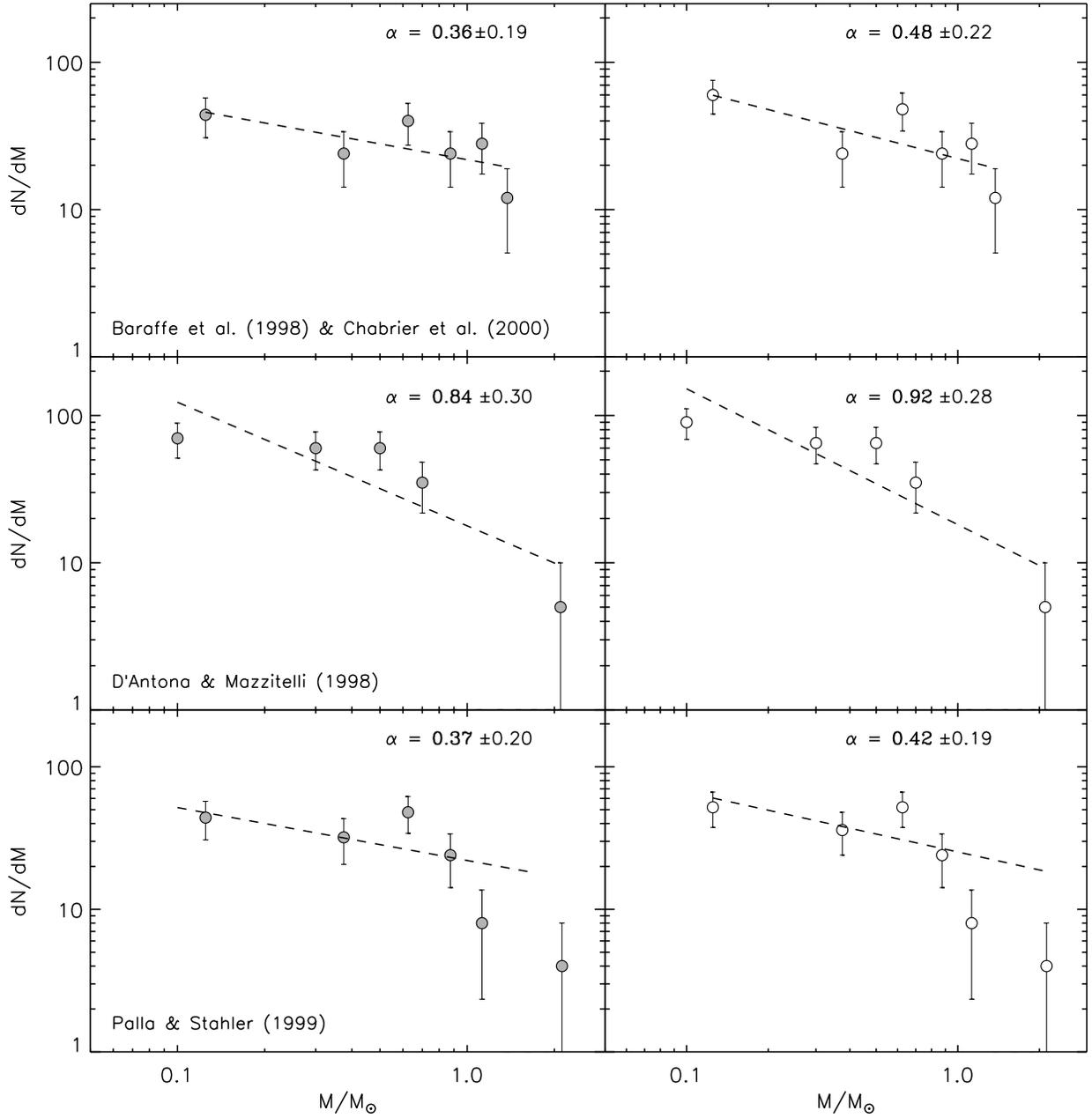}
\caption{The plots of $\frac{dN}{dM}$ versus mass of the certified PMS
      objects in Cha~II, depending on the indicated evolutionary
      tracks, are shown in the left panels. The same plots when
      including the candidates are shown in the right panels.
      The fits to the points are represented as dashed lines and
      the slope of the fits are indicated.
       \label{chaii_imf}}
\end{figure}

As mentioned in previous sections, recent investigations and this
work provide a PMS population in Cha~II which is basically complete.
Although some of the candidates still lack spectroscopic confirmation,
this census is useful for a first inspection on the mass spectrum of
the region. By assuming that all the candidates are true Cha~II members,
we attempted the IMF determination by using the approximation
for the low-mass end of the mass function, $ \frac{dN}{dM} \propto M^{-\alpha}$
\citep{Mor03}, adopting mass bins of 0.2~$M_{\odot}$ (c.f. Figure~\ref{chaii_imf});
this value is larger than the accuracy on mass estimates derived from the 
uncertainties on temperature and luminosity and, at the same time, allows 
to have a statistically valid number of objects at each bin.
The value of the slope $\alpha$ varies between 0.4 and 1 depending on the 
adopted evolutionary tracks. The resulting average value in the mass 
range $0.1\lesssim M \lesssim 1$~$M_{\odot}$ is $\alpha$=0.5-0.6 depending 
on the inclusion of the candidates (see Figure~\ref{chaii_imf}).

Despite the problems arising mainly from the still uncertain nature
of some of the candidates, as well as from the presence of 
unresolved binaries, our estimate of the IMF slope in 
Cha~II is fairly consistent with the IMF slopes measured 
in other the T associations \citep[c.f.][]{Lop04, Com00, Bri02} 
and also matches those derived for several star clusters under very
different age and formation history conditions
\citep[c.f.][]{Bar98b, Mor03, Bej01, Luh00, Tej02}\footnote{We stress
that in order to draw this conclusion, we have homogeneously 
determined the slope of the IMF in the other star forming regions 
using the same sets of evolutionary tracks.}. Thus, notwithstanding
the uncertainties and low statistics, our first IMF study in Cha~II
gives further evidence of the stellar IMF being roughly invariant
down to the Hydrogen burning limit.

The binary star fraction (BSF) in Cha~II is expected to be on the order 
of 13-15\% \citep{Bra96, Koh01, Alc07} and does not significantly affect 
the shape of the IMF. By assuming a BSF=15\% and a uniform distribution 
of this fraction of unresolved binaries in the range 0.1-2 $M_{\odot}$, 
we used the simulation of binary systems by \citet{Pre99} to investigate 
their effects on the stellar positions in the HR diagram and hence, on 
the IMF. We estimate a variation of the $\alpha$ slope to be $\sim$0.1, 
i.e. well within the uncertainty due to modeling.

Four of the seven objects (i.e. $\sim$ 60\%) below the H-burning
limit are not yet spectroscopically confirmed. Therefore, any
concluding statement on the sub-stellar IMF in Cha~II is still
premature and must be deferred to the future, when the proposed
sub-stellar candidates will be confirmed or rejected as members
of the cloud. If all the sub-stellar candidates were confirmed
as members, then the sub-stellar IMF in Cha~II would be as in
other T associations, otherwise it would have a significant 
drop, with only three sub-stellar members.

\subsection{Star Formation Efficiency and Star Formation Rate}
\label{SFE}

Average values of stellar masses and ages have already been used in Paper~I
to draw some conclusions on the Star Formation Efficiency (SFE) and
Star Formation Rate (SF~rate) in Cha~II. In this section we investigate
the global SFE and SF~rate, but using the individual masses and ages of
the members as estimated in \S~\ref{sec_mass_age}. 
Note, however, that the stellar mass could not be estimated
for the five bright IRAS sources (see \S~\ref{sec_hrd}) 
and for the two objects classified as ``Continuum-type'', i.e. IRAS~F12571-7657 
and SSTc2d~J130521.7-773810 (see \S~\ref{SpecTyp}). 
Moreover SSTc2d~J130529.0-774140 and
IRAS~13036-7644 are not detected in optical and near-IR wavelengths
and hence, their stellar parameters could not be determined.
In performing the following calculations, we assumed for these nine 
objects the average mass, i.e. 0.5~$M_{\odot}$. However, their 
inclusion does not affect the final results.

We derived the SFE as in Paper~I, i.e. $SFE=\frac{M_{star}}{M_{cloud}+M_{star}}$
where $M_{cloud}$ is the cloud mass, and $M_{star}$ is the total mass in PMS 
objects. We use the estimate of the cloud mass (670~M$_{\odot}$, see Paper~I)
as derived from the c2d extinction map with the 240 arc-sec resolution;
other estimates are present in the literature which vary
between 700 and 1250 M$_{\odot}$ depending on the tracer molecule used.
Considering the uncertainty on the cloud mass and on the total mass in PMS
objects, which depends on the adopted evolutionary tracks (Table~\ref{tab:SF}), 
we find that the SFE in Cha~II ranges between 1-4\%, regardless of the 
inclusion of the candidates. Lower values (1-2\%) are found when considering 
the YSOs alone. A summary of the global SFE estimates is reported in 
Table~\ref{tab:SF}. Our SFE results agree with previous estimates
\citep[$\sim$1\%,][]{Miz99} and match the typical value of a few per cent found
for T associations \citep{Miz95, Tac96, Miz99}. A considerably higher value for
the Cha~II cloud mass (i.e. 1860~M$_{\odot}$) was reported by \citet{Miz01} on
the basis of a $^{12}CO$ map of the cloud. Using this estimate, we obtain values
around 1\% for the SFE, which are still consistent with the range of values
expected in T associations. Interestingly, the SFE in Cha~II is lower than
in Cha~I \citep[7-13\%,][]{Miz99} and this may be a consequence of a different
history of star formation in the Chamaeleon clouds, as discussed in Paper~I.

The SF~rate has been determined using the total mass of PMS objects
and candidates in Cha~II and their average age as reported in
Table~\ref{tab:SF}. We performed this exercise by using three
different sets of evolutionary models and the results are summarized in
Table~\ref{tab:SF}. Considering the whole sample of PMS objects and
candidates, we find that the Cha~II cloud is turning some 8$\pm$2~$M_{\odot}$
into stars every Myr. The error depends on the uncertainty of the
evolutionary models.

\section{Disks vs. stellar properties}
\label{star_disk}
 
In this section we report possible links between the disk parameters and 
stellar properties. We use the disk parameters determined in Paper~I, but 
we caution the reader that the relationships we find may be rather noisy
because, despite the completeness of the sample, the number of sources in 
Cha~II is low. Furthermore, disk masses are poorly constrained given the 
lack of data at mm wavelengths for the majority of the sources. 
In this sense, the results we discuss here should be considered 
as preliminary.

\begin{figure}[h]
\epsscale{0.9}
\plotone{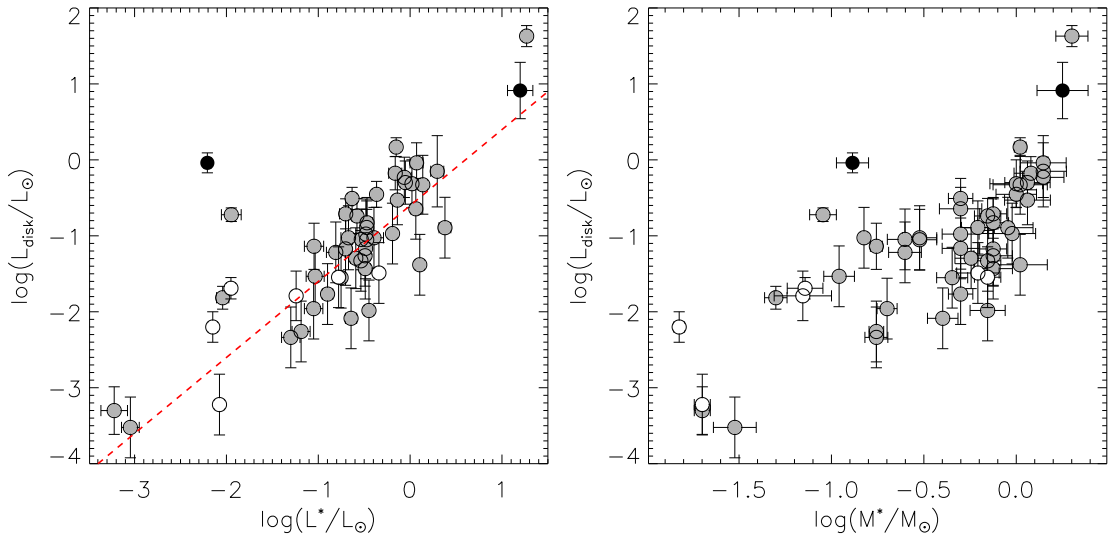}
\caption{Disk luminosity as function of stellar luminosity (left panel) 
and disk luminosity as a function of stellar mass (right panel) 
for the PMS stars and candidates in Cha~II.
The dashed line represents the $L_{disk}/L_\star = 0.25$ relationship 
as estimated by \citet{All06a}. Symbols are as in Figure~\ref{hrd}. 
\label{Ldisk_Lstar}}
\end{figure}


Figure~\ref{Ldisk_Lstar} (left panel) shows the disk luminosity 
as a function of stellar luminosity; a trend for the disk luminosity 
to increase with the stellar luminosity is observed across more 
than 3 orders of magnitude in stellar luminosity. Though the 
relation is rather scattered, it seems to follow the relationship 
$L_{disk}/L_\star = 0.25$ (dashed line in Figure~\ref{Ldisk_Lstar}),  
which can be interpreted as the luminosity of the disk deriving from 
reprocessed stellar light under the assumption of a flat, optically 
thick disk extending outward from the stellar surface \citep{All06a}. 
Note that the class I object IRAS~12500-7658 (black dot at  
$\log L_{disk} \approx 0$ in Figure~\ref{Ldisk_Lstar}) was not observed 
in our follow-up spectroscopy. Its stellar parameters are then uncertain 
and, in particular, its stellar luminosity might be quite underestimated.

The PMS mass-luminosity relation implies a secondary correlation between 
the stellar mass and disk luminosity. Such correlation (Figure~\ref{Ldisk_Lstar}, 
right panel) seems to extend to the very low mass regime. Thus, more massive 
stars possess more luminous disks and viceversa.

In Paper~I the disk mass for several of the studied objects was estimated 
by modeling the SEDs with appropriate grids of circumstellar envelope/disk 
models \citep{Dul01, DAl05, Rob06}. For objects with masses greater than 
$\sim$0.1~$M_{\odot}$ there seems to be a very weak indication that more 
massive stars may have more massive disks. However, the relationship is 
very noisy. Because of the lack of millimeter data, the disk masses are 
not really well constrained, which may explain the large 
scatter in the possible correlation between 
disk mass and stellar mass.

Another study \citep{Car02} found that, despite the similar stellar 
ages and conditions (i.e. no presence of high ultraviolet radiation fields), 
PMS stars in the IC~348 cluster possess less massive disks than those 
in Taurus \citep{Bec90, Ski91, Ost95, Hen98}. In numbers, 14\% of the
PMS stars in Taurus with a mass M$_\star >$0.27~M$_\odot$ have disks
more massive than 0.025~M$_\odot$, while none exceeds this limit in
IC~348. Also in the Orion Nebular Cluster, \citet{Mun95} exclude the presence 
of massive disks \citep[see also][and references therein]{Wil06}, while 
submillimter data obtained by \citet{And07} show that disk properties of the 
young population in the $\rho$~Ophiuchus dark cloud match those observed 
in Taurus. This difference in disk masses might be a consequence of the 
``clustered'' star-forming mode with respect to the ``isolated'' one. 
However, this is not yet evident, given the limited sample of star forming 
regions for which millimeter continuum surveys are available.
The disk masses given in Paper~I are based on disk models and, hence, 
are highly uncertain. However, at first approximation, some 10\% of the 
objects with M$_\star >$0.27~M$_\odot$ have M$_{disk} >$0.025~M$_\odot$, 
i.e. a percentage similar to that measured in Taurus.


\section{H$\alpha$ as accretion indicator}
\label{Ha_accr_ind}
 
\begin{figure}[h]
\epsscale{0.7}
\plotone{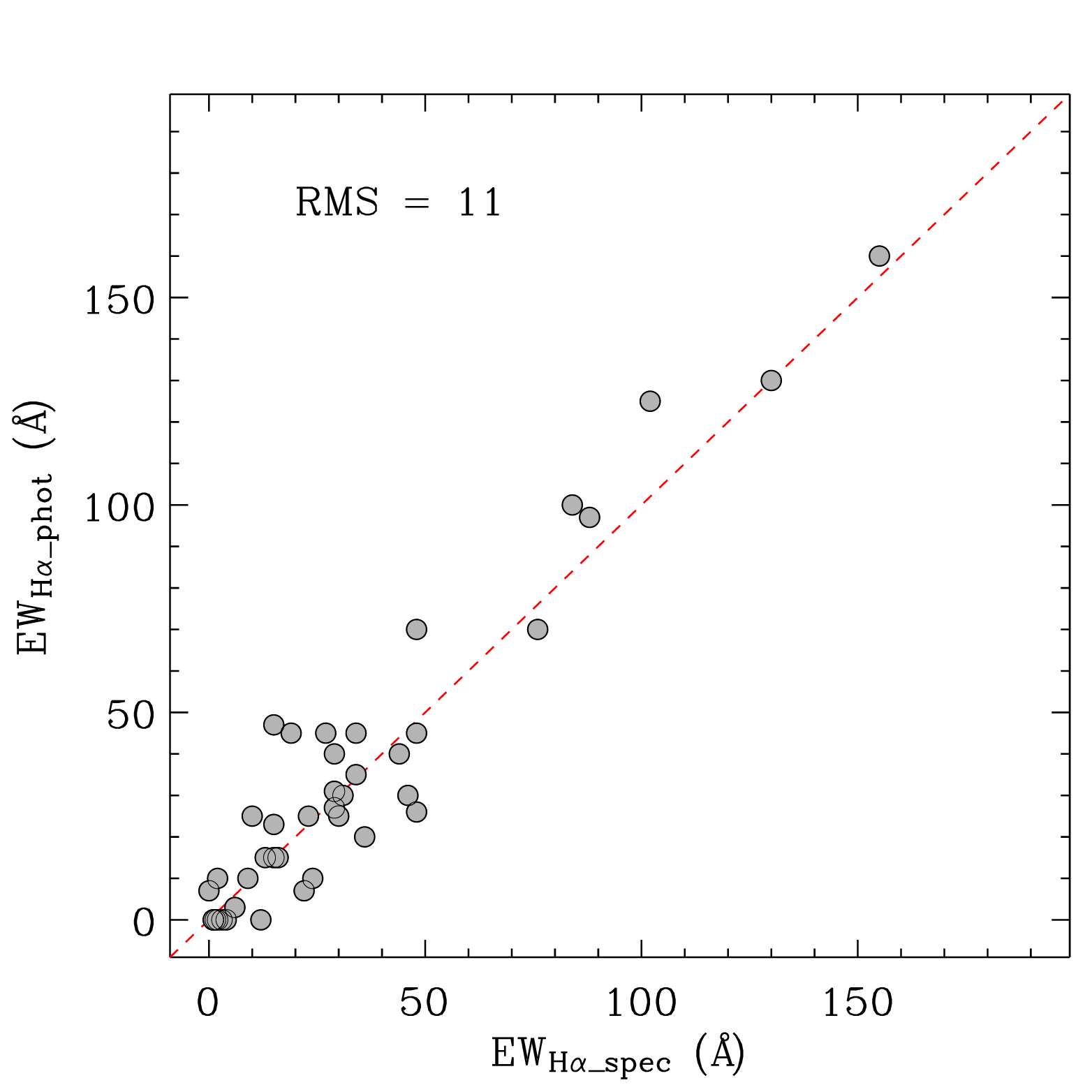}
\caption{Comparison between the H$\alpha$ equivalent width derived from
our spectroscopy (EW$_{\rm H\alpha\_spec}$) and that obtained from the
WFI ($H\alpha_{12}-H\alpha_7$) color index (EW$_{\rm H\alpha\_phot}$)
determined as described in \citet{Spe07}.
\label{comp_WHa}}
\end{figure}

The presence of a strong H$\alpha$ emission line in young objects is
interpreted as a signature of the accretion process \citep[][and references therein]{Fei99}.
In this section we investigate the H$\alpha$ properties of the Cha~II stars
in relation with other accretion and disk indicators derived on the basis of
the Spitzer data reported in Paper~I.
The EW$_{H\alpha}$ was measured on the spectra of the confirmed PMS objects in Cha~II, 
while for the candidates we used the WFI ($H\alpha_{12}-H\alpha_7$) color index and 
the calibration relation between this index and the EW$_{H\alpha}$ computed 
by \citet{Spe07}. As shown in Figure~\ref{comp_WHa}, this method yields results 
consistent with those obtained from spectroscopy when applied to the confirmed 
PMS stars within $\sim$10~\AA.

In Paper~I we used the spectral index $\alpha_{[K \& {\rm MIPS1}]}$ to
investigate the different object classes according to the \citet{Lad06}
criteria. According to our definitions in Paper~I, we found that some
70\% of the PMS objects in Cha~II are Class II sources, i.e. have a SED
with spectral index $\alpha_{[K \& {\rm MIPS1}]}>-$1.6. 
The dividing line between accretion and chromospheric H$\alpha$ emission 
varies with spectral type \citep{Whi03, Lad06}. Using the definition by \citet{Whi03} 
to divide weak from classical T Tauri stars, we find that 72\% of the objects 
in Cha~II would be classified as classical, in perfect agreement with the spectral index 
$\alpha_{[K \& {\rm MIPS1}]}$ criterion.

\begin{figure}[h]
\epsscale{0.7}
\plotone{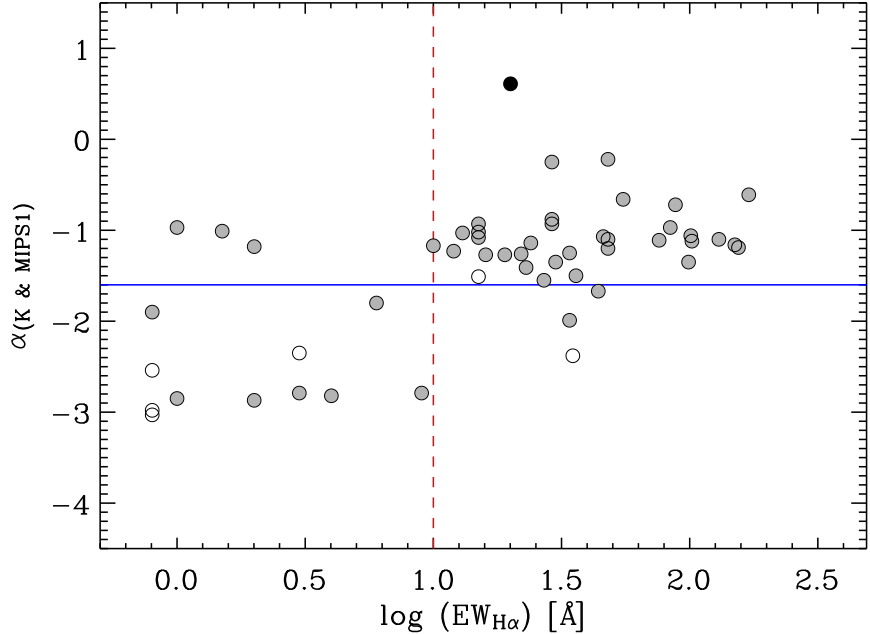}
\caption{H$\alpha$ equivalent width as a function of the
$\alpha_{[K \& {\rm MIPS1}]}$ spectral index.
The horizontal continuous line
marks the $\alpha_{[K \& {\rm MIPS1}]}$ limit for optically thick
disks, while the dashed line represents the EW$_{H\alpha}$ limit
for accretion emission. Symbols are as in Figure~\ref{hrd}.
\label{WHa_a}}
\end{figure}

Figure~\ref{WHa_a} shows the H$\alpha$ Equivalent Width (EW$_{H\alpha}$)
as a function of $\alpha_{[K \& {\rm MIPS1}]}$ for the PMS stars
and candidates in Cha~II. From this figure we find that more than 80\%
of the Class II sources (i.e. $\alpha_{[K \& {\rm MIPS1}]}$
indicative of optically thick disk) have EW$_{H\alpha}$
larger than 10~\AA, while 75\% of the Class III sources
(i.e. $\alpha_{[K \& {\rm MIPS1}]}$ indicative of thin or no disk)
have weaker H$\alpha$ emission. 
Here we use EW$_{H\alpha}$=10~\AA~ as the approximate lower limit for accretion. 
Because the Cha~II population mainly consists of late K and early 
M-type objects, in agreement with the definition by \citet{Whi03}.

The trend in Figure~\ref{WHa_a} is consistent with that found in other
regions \citep[e.g. IC~348,][]{Lad06} and may suggest that the gaseous 
and dust components of disks evolve on similar timescales. 
The three objects with  $\alpha_{[K \& {\rm MIPS1}]}<$--1.6 but EW$_{H\alpha}>$10\AA~
are IRAS~12448-7650, Sz~47 and Sz~60W. IRAS~12448-7650 
was not observed in our follow-up spectroscopy; its T$_{\rm eff}$ and EW$_{H\alpha}>$10\AA~ 
are estimated from the WFI color indexes and hence 
are rather uncertain. Sz~47 and Sz~60W are confirmed cloud members, though 
the stellar parameters of Sz~47 are very uncertain (\S~\ref{SpecTyp}).
For both these objects we have signatures of accretion 
(see later in this paragraph and Table~\ref{pms_objects_a})
and they might then be cases where the disk has evolved 
but still substantial gas is present and accreting
onto the central star.

\begin{figure}[h]
\epsscale{0.7}
\plotone{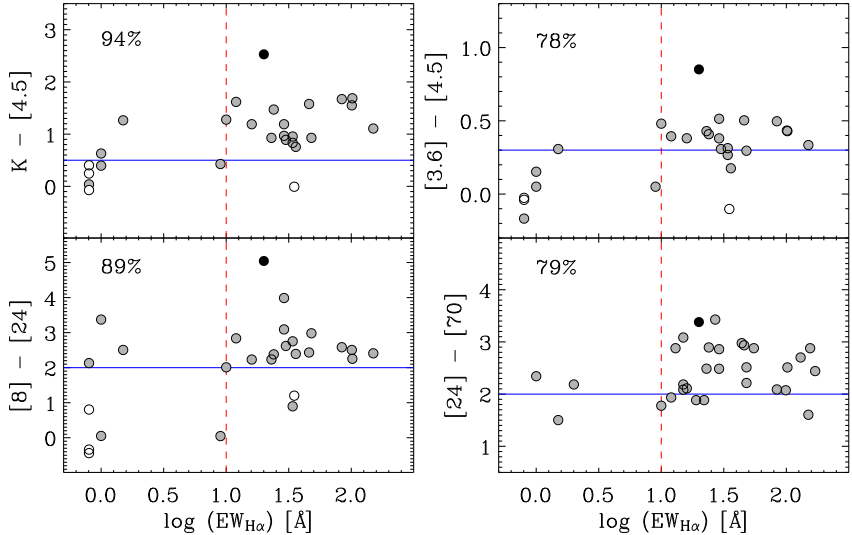}
\caption{H$\alpha$ equivalent width as a function of different
Spitzer colors for PMS objects and candidates in Cha~II.
The horizontal continuous lines mark the minimum color excess
for T Tauri stars as estimated by \citet{Rob06}, for
$K-[4.5]$, $[3.6]-[4.5]$ and $[8]-[24]$, and by \citet{You05}
for $[24]-[70]$. The dashed lines represent the EW$_{H\alpha}$
limit for accretion emission. 
The percentage of objects with EW$_{H\alpha}$ above this 
limit and whose infrared color exceeds the threshold 
fixed for T Tauri stars is indicated in each panel 
(see text). Symbols are as in Figure~\ref{hrd}. \label{WHa_SpitCol}}
\end{figure}

\begin{figure}[h]
\epsscale{0.5}
\plotone{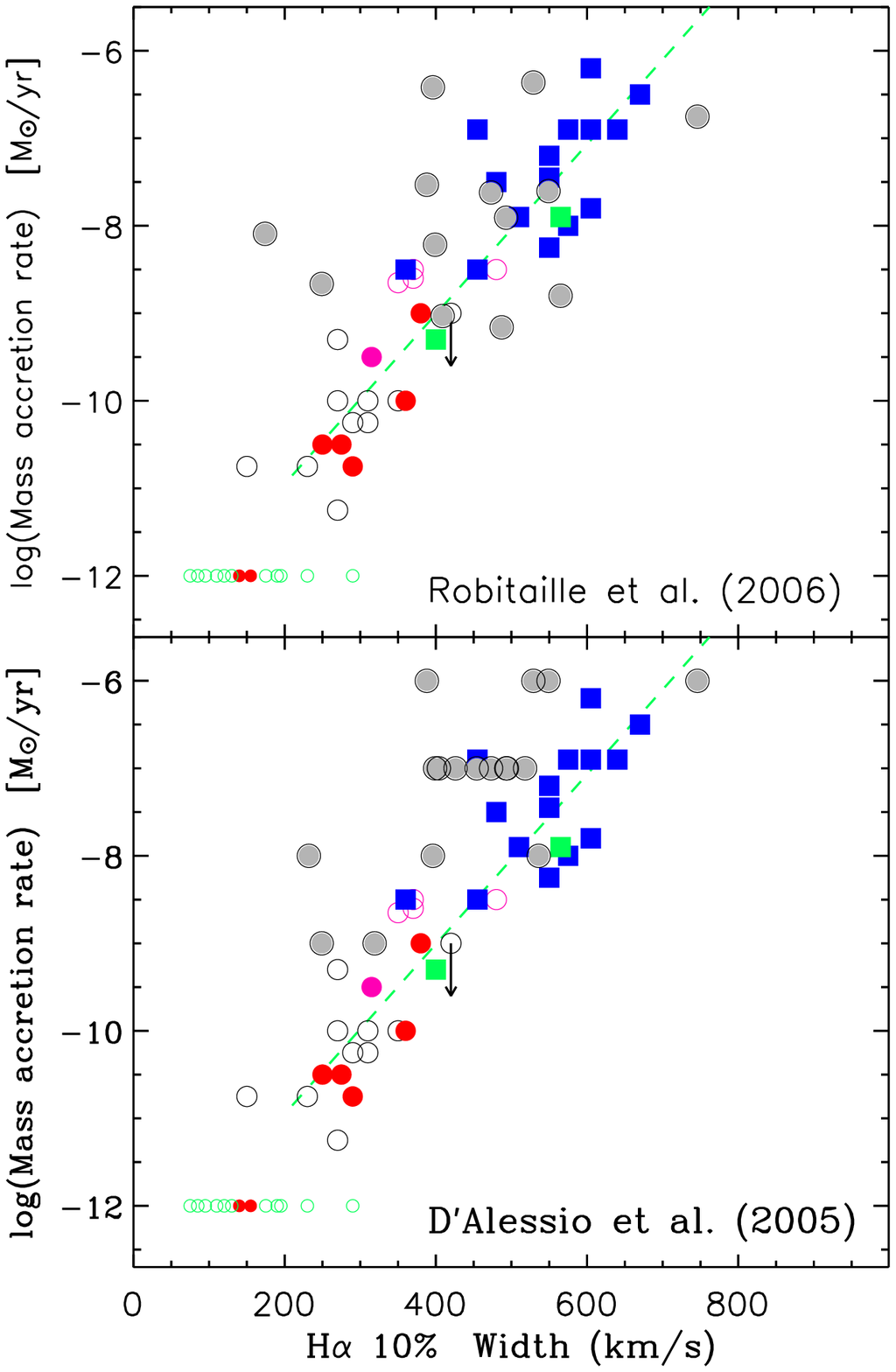}
\caption{Mass accretion rate as a function of the H$\alpha$
full width at 10\% peak intensity for PMS objects adapted
from \citet{Nat04}; the dashed line represents the calibration
relation derived by these authors. The grey dots represent the
Cha~II PMS members. \label{Ha10_Maccr} }
\end{figure}


Being responsible for the IR excess emission, the presence of a dust disk can be
traced by $\alpha_{[K \& {\rm MIPS1}]}$, i.e. a measure of the ($K$-24$\mu$m)
color excess, as well as a number of other IR colors. \citet{Har98} and \citet{Lad06}
have already shown that stars with strong H$\alpha$ emission in Taurus and IC~348
tend to possess large ($K-L$) excess, or the equivalent ($K-4.5\mu m$) excess when
using the Spitzer bands. We investigated the relation between the EW$_{H\alpha}$
and a number of Spitzer colors for the PMS stars in Cha~II in order to identify
possible tracers of the accretion process.

In the upper-left panel of Figure~\ref{WHa_SpitCol}, the $K-[4.5]$ color versus
EW$_{H\alpha}$ relation is shown; we found that more than 90\% of the objects
with EW$_{H\alpha}$ above the approximate limit for accretion (i.e. 10~\AA)
also have $K-[4.5]>$0.5, i.e. the minimum excess for Stage~II objects
(i.e. classical T Tauri stars) according to the models by \citet{Rob06}.
Despite the low number of weak-line T Tauri stars in Cha~II,
some 70\% of the objects with no prominent H$\alpha$ emission have
$K-[4.5]$ color falling below this limit.
Thus, consistently with the results by \citet{Har98} and \citet{Lad06}, we conclude
that the $K-[4.5]$ color index is a useful diagnostic for accretion, with
0.5~mag the approximate threshold.
Among the other color indexes inspected in the wavelength range covered by
Spitzer (i.e. 3.6-70~$\mu$m), we found that $[3.6]-[4.5]$, $[8]-[24]$ and $[24]-[70]$
also tend to be larger for objects with stronger H$\alpha$ emission.
The correlation between these colors and the EW$_{H\alpha}$ is
shown in the other panels of Figure~\ref{WHa_SpitCol}.
As above, we consider the minimum color excesses, estimated by \citet{Rob06} and
\citet{You05} for Stage~II objects, as approximate threshold for the accretion
process. We found that the percentage of objects with EW$_{H\alpha} >$10~\AA~ and
color excess above the fixed threshold varies between 80\% and 90\% depending on
the considered color, while some 50-80\% of the objects with weaker H$\alpha$
emission have smaller colors. Since the emission at longer wavelengths 
is coming from farther out in the disk, the longer wavelength colors probably 
trace disk flaring, which may trace younger disks, which at the same time
are more likely to be the high accretors.
Because of the large scatter in the correlation between
the EW$_{H\alpha}$ and and the investigated 
infrared color indices (Figure~\ref{WHa_SpitCol}), 
these indices can be used as overall indicators of accretion, 
but cannot replace detailed accretion studies in individual objects.

Another way to investigate the accretion diagnostics uses the width of
Balmer emission lines. \citet{Nat04} have shown that the observed width of
the H$\alpha$ line can be used not only to discriminate between accretors
and non-accretors, but also to get an approximate estimate of the mass
accretion rate ($\dot{M}$). Using our spectra and following the prescription
by \citet{Nat04}, we determined the full width of H$\alpha$ measured at 10\% of
the peak intensity (H$\alpha$~10\%), after continuum subtraction, for the
PMS objects in Cha~II. Although our GIRAFFE-MEDUSA spectra have a resolution four
times less than the one used by \citet{Nat04}, we can measure the width of the
H$\alpha$ line with an accuracy of some 10~Km/s.
Figure~\ref{Ha10_Maccr} is a reproduction of Figure~3 by \citet{Nat04}.
The dashed line represents the  calibration relation determined by these
authors. The grey dots represent the Cha~II members; the H$\alpha$~10\% 
was computed using our FLAMES spectra, as described above, while $\dot{M}$
was derived by the fits of the disk accretion models by \citet{Rob06} and \citet{DAl05}
to the SEDs as reported in Paper~I. A fair agreement with the \citet{Nat04}
data is found, but some objects show a larger scatter, possibly due to
the lower resolution of our spectra with respect to those used by \citet{Nat04}. 
The H$\alpha$~10\% for the PMS objects in Cha~II and their $\dot{M}$ as derived 
by using the calibration relation by \citet{Nat04} are reported in Table~\ref{pms_objects_a}.



\section{Summary}
\label{sum}

We presented the spectroscopic follow-up of PMS objects and candidates recently 
selected in the Cha~II dark cloud on the basis of the Spitzer c2d Legacy survey 
and complementary optical and near-IR data \citep{Alc07}.
We confirmed the PMS nature for 51 objects, which display the typical
spectral signatures of young low-mass stars, i.e. Lithium absorption,
H$\alpha$ emission and a late-type spectrum, while 11 candidates
still need spectroscopic investigation in order to assess their PMS
nature. More than 96\% of the observed c2d YSO candidates resulted 
in true Cha~II members, supporting the reliability of the c2d selection
criteria \citep{Har07, Eva07}.

The stellar parameters of the PMS objects and candidates have
been determined. Most objects in Cha~II have masses in the range
$0.015 \leq M \leq 1 M_{\odot}$. The average mass is 0.5$\pm$0.1~$M_{\odot}$,
with the error representing the uncertainty due to modeling. The mean
age of the Cha~II population ranges from 2 to 6~Myr depending on the
adopted model. The average age is 4$\pm$2~Myr with the error coming
from modeling alone. The average age drops to about 2~Myr when considering
the YSOs, i.e. the IR-excess population selected from the c2d multi-color
criteria. Seven objects (three confirmed plus four candidates)
are sub-stellar and another five are confirmed to be very low-mass PMS
stars close to the Hydrogen burning limit. In addition, the low 
luminosity ($L/L_\odot \approx$ 0.007) object detected in the optical
and matching the position of the Class~I source IRAS12500-7658 might also 
have a mass close to the sub-stellar limit or may be a highly
extincted object.

Results on the global properties of star formation in Cha~II are 
the following. The average value of the IMF slope {bf ($\alpha$=0.5-0.6)} 
in the mass range $0.1\lesssim M \lesssim 1$~$M_{\odot}$ is fairly 
consistent with those measured in other T associations. 
Our first study of the mass spectrum
in Cha~II seems to provide further evidence for the stellar IMF being
roughly the same as the field IMF down to the Hydrogen burning limit, 
but the results depend on the adopted evolutionary models.

The star formation efficiency of 1-4\% is similar
to our estimates for other T associations like Taurus and Lupus,
but significantly lower than for Cha~I ($\sim$7\%). This suggests that
different star-formation activities in the Chamaeleon clouds may reflect
a different history of star formation.
The Cha~II cloud turns about 8$M_{\odot}$ into stars every Myr,
i.e. much less than the star formation rate in the other c2d clouds.
However, the star formation rate is not steady and there is evidence
of acceleration of star formation, as seen in other star forming
regions. This, together with the results of Paper~I on the high disk
fraction, may indicate that the star formation in Cha~II possibly
occurred rapidly a few million years ago, which would explain
the scarcity of disk-less stars in the region.

We have also investigated possible links between the stellar and the
disk properties. We find a fair indication that more luminous objects 
have more luminous disks, which suggests a possible correlation between 
stellar mass and disk luminosity via the PMS mass-luminosity relation. 
About 10\% of the objects with M$_\star >$0.27~M$_\odot$ possess disks 
more massive than 0.025~M$_\odot$, further supporting recent ideas that 
the presence of massive disks might be a consequence of the ``isolated'' 
star-forming mode with respect to the ``clustered'' one.

Finally, the strength of the H$\alpha$ emission line of the PMS stars
in Cha~II behaves as in other young populations in other regions.
We find that objects with optically thick disks show strong H$\alpha$
emission, suggesting that the gaseous and dust components of disks
evolve on similar timescales. The correlation between the H$\alpha$
emission and the IR excess emission allowed us to identify various
Spitzer colors in the range 3.6-70~$\mu$m that can be used
as tracers of the accretion process, but different colors trace
different disk regions.

\acknowledgments

We are grateful to anonymous referee for his/her very constructive 
comments and suggestions.
This paper is based on observations carried out at the European Southern Observatory,
Paranal (Chile), under observing programs numbers 076.C-0385 and 078.C-0293.
This work was partially financed by the Istituto Nazionale di Astrofisica (INAF)
through PRIN-INAF-2005.
Support for this work, part of the Spitzer Legacy Science Program, was
provided by NASA through contract 1224608 issued by the Jet Propulsion
Laboratory, California Institute of Technology, under NASA contract 1407.
L.S. acknowledges financial support from PRIN-INAF-2005 (Stellar clusters:
a benchmark for star formation and stellar evolution). 
We are grateful to Dan Jaffe for his careful reading and for his valuable 
comments/suggestions to the paper.
We thank the c2d collaborators for the many discussions and suggestions
during the telecons.
This publication makes use of data products from the Two Micron All Sky
Survey, which is a joint project of the University of Massachusetts and
the Infrared Processing and Analysis Center/California Institute of
Technology, funded by NASA and the National Science Foundation. We also
acknowledge extensive use of the SIMBAD data base.
We are also grateful to many others, in particular to Salvatore Spezzi.

{}


\clearpage


\begin{deluxetable}{lllllllllll}
\tabletypesize{\tiny}

\tablecaption{\label{pms_objects_a} 
Spectral types and line widths for pre-main 
sequence objects and candidates in Cha~II. }

\tablewidth{0pt}

\rotate

\tablehead{
\colhead{No.} & \colhead{Object} &  \colhead{FLAMES$^\Theta$} &  \colhead{No.} & \colhead{S/N} &
\colhead{PMS} & \colhead{SpT$^\Psi$} & \colhead{EW$_{\rm H\alpha}^\xi$} & \colhead{EW$_{\rm LiI}$} & \colhead{EW$_{\rm H\alpha}$~10\%} & \colhead{$\dot{M}$}\\ 
 & \colhead{id.} & &  \colhead{Spec.} &  &
\colhead{Flag$^\Upsilon$} &  & [\AA] & [\AA] &[Km/s] & [M$_{\odot}$/yr] 
}

\startdata

1   &   IRAS~12416-7703 / AX~Cha                   & --   & -- &   --  & CND	      &  M2.5   &   15    &  --  &  --      &      --    \\  
2   &   IRAS~12448-7650                            & --   & -- &   --  & CND	      &  M0.5   &   35    &  --  &  --      &      --    \\  
3   &   IRAS~F12488-7658 / C~13                    & --   & -- &   --  & CND	      &  M5.5   &    0    &  --  &  --      &      --    \\  
4   &   IRAS~12496-7650 / DK~Cha                   & GM   & 2  &   45  & PMS	      &  F0     &   88    &  0   &   746    &   2.2e-06  \\  
5   &   WFI~J12533662-7706393                      & --   & -- &   --  & CND	      &  M6     &    --   &  --  &  --      &      --    \\  
6   &   C~17                                       & --   & -- &   --  & CND	      &  M1.5   &    0    &  --  &  --      &      --    \\  
7   &   IRAS~12500-7658$^\bullet$                  & --   & -- &   --  & PMS	      &  M6.5   &   20    &  --  &  --      &      --    \\  
8   &   C~33                                       & --   & -- &   --  & CND	      &  M1     &    0    &  --  &  --      &      --    \\  
9   &   IRAS~12522-7640  / HH~54                   & --   & -- &   --  &	      &  --     &    --   &  --  &  --      &      --    \\  
10  &   Sz~46N                                     & U    & 2  &   20  & PMS	      &  M1     &   16    & 0.58 &   232    &   2.2e-11  \\  
11  &   Sz~47                                      & GM   & 3  &  300  & PMS	      &  --     &   34    &  0   &   845    &   2.0e-05  \\  
12  &   IRAS~12535-7623 / CHIIXR~2                 & U    & 2  &   60  & PMS$^\dag$   &  M0     &   15    & 0.57 &   396    &   8.9e-10  \\	
13  &   SSTc2d~J125758.7-770120 / \#1 $^\Gamma$    & --   & -- &   --  & PMS	      &  M9     &    --   &  --  &  --      &      --    \\  
14  &   ISO-CHAII 13$^\Omega$                      & --   & -- &   --  & PMS	      &  M7     &   101   &  --  &   283    &   7.1e-11  \\  
15  &   WFI~J12583675-7704065                      & --   & -- &   --  & CND	      &  M9     &    --   &  --  &  --      &      --    \\  
16  &   WFI~J12585611-7630105                      & GM   & 3  &   50  & PMS$^\dag$   &  M5     &   31    & 0.50 &   388    &   7.4e-10  \\  
17  &   IRAS~12553-7651 / ISO-CHAII 28$^\bullet$   & --   & -- &   --  & PMS	      &  K4.5   &    --   &  --  &  --      &      --    \\  
18  &   C~41$^\Delta$                              & GM   & 1  &   15  & PMS	      &  M5.5   &   48    & 0.36 &   487    &   6.8e-09  \\  
19  &   ISO-CHAII 29                               & GM   & 2  &   60  & PMS$^\dag$   &  M0     &   1     & 0.52 &   249    &   3.3e-11  \\  
20  &   IRAS~12556-7731                            & GM   & 2  &  100  & PMS$^\dag$   &  M5     &   0     & 0.48 &   0      &   1.2e-13  \\  
21  &   WFI~J13005297-7709478                      & --   & -- &   --  & CND	      &  M9     &    --   &  --  &  --      &      --    \\  
22  &   Sz~48NE / CHIIXR~7$^\Sigma$                & --   & -- &   --  & PMS	      &  M0.5   &   22    &  --  &  --      &      --    \\  
23  &   Sz~49 / ISO-CHAII 55 / CHIIXR~9 $^\Sigma$  & --   & -- &   --  & PMS	      &  M0.5   &   170   &  --  &  --      &      --    \\  
24  &   Sz~48SW / CHIIXR~7                         & GM   & 3  &   90  & PMS$^\ddag$  &  M1     &   19    & 0.61 &   445    &   2.6e-09  \\  
25  &   Sz~50 / ISO-CHAII 52 / CHIIXR~8            & U    & 2  &   25  & PMS	      &  M3     &   29    & 0.49 &   265    &   4.7e-11  \\  
26  &   WFI~J13005531-7708295                      & GM   & 2  &  100  & PMS$^\dag$   &  M2.5   &   2     & 0.55 &   254    &   3.7e-11  \\  
27  &   RXJ1301.0-7654a                            & U    & 1  &   60  & PMS	      &  K5     &   4     & 0.58 &   --     &      --    \\  
28  &   IRAS~F12571-7657 / ISO-CHAII 54            & GM   & 3  &   25  & PMS$^\dag$   &  C      &   12    & 0.39 &   529    &   1.7e-08  \\  
29  &   Sz~51                                      & U    & 2  &   50  & PMS	      &  K8.5   &   102   & 0.35 &   399    &   9.5e-10  \\  
30  &   CM~Cha / IRAS~12584-7621                   & GM   & 1  &  300  & PMS$^\ddag$  &  K7     &   29    & 0.38 &   388    &   7.4e-10  \\  
31  &   C~50                                       & GM   & 2  &   20  & PMS$^\dag$   &  M5     &   36    & 0.46 &   457    &   3.4e-09  \\  
32  &   IRAS~12589-7646 / ISO-CHAII 89             & --   & -- &   --  & CND	      &  M4     &   3     &  --  &   --     &      --    \\  
33  &   RXJ1303.1-7706                             & GM   & 3  &  300  & PMS	      &  M0     &   3     & 0.60 &   179    &   7.0e-12  \\  
34  &   C~51                                       & GM   & 2  &   35  & PMS$^\dag$   &  M4.5   &   9     & 0.50 &   256    &   3.9e-11  \\  
35  &   WFI~J13031615-7629381                      & --   & -- &   --  & CND	      &  M7     &   15    &  --  &   --     &      --    \\  
36  &   Hn~22 / IRAS~13005-7633 $^\Lambda$         & --   & -- &   --  & PMS	      &  M2     &   55    &  --  &   --     &      --    \\  
37  &   Hn~23                                      & U    & 1  &  110  & PMS$^\ddag$  &  K5     &   13    & 0.52 &   --     &      --    \\    
38  &   Sz~52                                      & GM   & 2  &   35  & PMS$^\ddag$  &  M2.5   &   48    & 0.39 &   454    &   3.2e-09  \\    
39  &   Hn~24                                      & GM/U & 3/2&100/55 & PMS$^\ddag$  &  M0     &   1     & 0.58 &   319    &   1.6e-10  \\	
40  &   Hn~25                                      & GM   & 1  &   50  & PMS$^\ddag$  &  M2.5   &   24    & 0.46 &   449    &   2.9e-09  \\	
41  &   Sz~53                                      & GM   & 2  &   80  & PMS	      &  M1     &   46    & 0.38 &   454    &   3.2e-09  \\  
42  &   Sz~54                                      & GM/U & 2/4&350/100& PMS	      &  K5     &   23    & 0.46 &   473    &   4.9e-09  \\  
43  &   SSTc2d~J130521.7-773810                    & GM   & 1  &   20  & PMS$^\dag$   &  C      &   29    & 0.44 &   404    &   1.0e-09  \\   
44  &   SSTc2d~J130529.0-774140                    & --   & -- &   --  & CND	      &  --     &    --   &   -- &  --      &      --    \\  
45  &   SSTc2d~J130540.8-773958 / \#5 $^\Gamma$    & --   & -- &   --  & PMS	      &  L1     &    --   &   -- &  --      &      --    \\  
46  &   Sz~55                                      & GM   & 3  &   90  & PMS	      &  M2     &  155    & 0.19 &   536    &   2.0e-08  \\  
47  &   Sz~56                                      & GM   & 4  &   15  & PMS	      &  M4     &   15    & 0.46 &   348    &   3.0e-10  \\  
48  &   Sz~57 / C~60                               & GM   & 3  &   75  & PMS	      &  M5     &   27    & 0.46 &   174    &   6.2e-12  \\  
49  &   Sz~58 / IRAS~13030-7707 /C~61              & GM   & 3  &  200  & PMS	      &  K5     &   15    & 0.46 &   493    &   7.8e-09  \\  
50  &   Sz~59                       		   & GM   & 3  &  200  & PMS	      &  K7     &   48    & 0.39 &   518    &   1.3e-08  \\  
51  &   C~62                       		   & GM   & 3  &   20  & PMS$^\dag$   &  M4.5   &   34    & 0.37 &   255    &   3.8e-11  \\  
52  &   Sz~60W                    		   & U    & 4  &   35  & PMS	      &  M1     &   44    & 0.51 &   426    &   1.7e-09  \\  
53  &   Sz~60E$^\Sigma$                 	   & --   & -- &   --  & PMS	      &  M4     &   99    &   -- &  --      &      --    \\  
54  &   IRAS~13036-7644 / BHR~86                   & --   & -- &   --  & PMS	      &  --     &    --   &   -- &  --      &      --    \\  
55  &   Hn~26                                      & GM   & 3  &  110  & PMS$^\ddag$  &  M2     &   10    & 0.59 &   494    &   7.9e-09  \\   
56  &   Sz~61                                      & GM   & 2  &  200  & PMS	      &  K5     &   84    & 0.38 &   549    &   2.7e-08  \\  
57  &   C~66                                       & GM   & 5  &   20  & PMS$^\dag$   &  M4.5   &   30    & 0.43 &   293    &   8.9e-11  \\   
58  &   IRAS~F13052-7653NW / CHIIXR~60             & GM   & 1  &  200  & PMS$^\dag$   &  M0.5   &   15    & 0.45 &   565    &   3.8e-08  \\   
59  &   IRAS~F13052-7653N  / CHIIXR~60             & U    & 1  &   40  & PMS$^\dag$   &  M1.5   &    2    & 0.58 &  --      &      --    \\   
60  &   Sz~62                                      & GM   & 2  &  150  & PMS	      &  M2.5   &   150   & 0.38 &   409    &   1.1e-09  \\  
61  &   Sz~63                                      & GM   & 1  &   90  & PMS	      &  M3     &   76    & 0.53 &   505    &   1.0e-08  \\  
62  &   2MASS13125238-7739182                      & GM   & 2  &   65  & PMS$^\dag$   &  M4.5   &   6     & 0.52 &   189    &   8.7e-12  \\	  
63  &   Sz~64                                      & GM   & 2  &   50  & PMS	      &  M5     &   130   & 0.40 &   241    &   2.8e-11  \\  

\enddata

\tablecomments{The coordinates are reported in Paper~I}
\tablenotetext{\Theta}{GM = GIRAFFE-MEDUSA ; U = UVES}
\tablenotetext{\Upsilon}{PMS = Pre-Main Sequence Object; CND = Candidate}
\tablenotetext{\Psi}{Spectral type of the candidate corresponding to 
the estimated temperature (see \S~\ref{Teff}). Spectral type 
equal to ``C'' stands for ``continuum-type object'' (see \S~\ref{SpecTyp})}
\tablenotetext{\xi}{The EW$_{\rm H\alpha}$ of candidates and objects lacking spectroscopy 
is estimated from the WFI ($H\alpha_{12}-H\alpha_7$) color index (see \S~\ref{Ha_accr_ind})}
\tablenotetext{\dag}{Cha~II member firstly confirmed in this work}
\tablenotetext{\ddag}{~Li{\sc i} absorption at $\lambda=$6707.8 \AA~ detected for the first time}
\tablenotetext{\Gamma}{Spectral type by \citet{Jay06}}
\tablenotetext{\Omega}{Spectral type and EW$_{\rm H\alpha}$ by \citet{Alc06}}
\tablenotetext{\Sigma}{Spectral type and EW$_{\rm H\alpha}$ by \citet{Hug92}}
\tablenotetext{\Lambda}{Spectral type by \citet{Har93}}
\tablenotetext{\Delta}{Given the low S/N of our FLAMES 
spectrum, we adopted for this object the spectral type by \citet{Bar04}}

\end{deluxetable}


\clearpage

\begin{table}
\begin{center}

\caption{Journal of FLAMES observations in the Cha~II dark cloud. \label{jour_obs} }

\begin{tabular}{ccccccc}

\tableline  
Field &   RA$^\dag$& DEC$^\dag$& Obs. Date   &  Exp. Time  &   No PMS            &  No   \\                
      &            &           &             &  (sec.)     &  (GM \& U)$^\ddagger$ & Field Stars   \\       
\tableline  
\tableline  

1     & 13:07:24.0 & -77:33:21 & 2006-03-01  &  3000   &   11+3      &  107  \\
1     &      "     &      "    & 2006-03-02  &  3300   &    7+3      &  100  \\
1     &      "     &      "    & 2006-03-03  &  3000   &    9+6      &  103  \\
2     & 13:02:54.2 & -77:18:03 & 2006-03-01  &  3000   &    3+1      &  116  \\
2     &      "     &      "    & 2006-03-02  &  3660   &    3        &   95  \\
2     &      "     &      "    & 2006-03-03  &  3000   &    4+1      &  121  \\
3     & 12:55:31.5 & -77:03:13 & 2006-03-01  &  3000   &    1        &  119  \\
3     &      "     &      "    & 2006-03-03  &  3000   &    1        &  120  \\
4     & 13:09:02.9 & -77:01:20 & 2006-03-01  &  3000   &    1+1      &  118  \\
5     & 13:02:06.5 & -76:48:37 & 2006-03-01  &  3000   &    2+2      &  118  \\
6     & 13:02:33.0 & -77:46:32 & 2006-03-01  &   900   &    5+2      &  111  \\
6     &      "     &     "     & 2006-03-03  &  3000   &    5+2      &  111  \\
7     & 13:11:20.6 & -77:49:30 & 2006-03-02  &  3000   &    5        &  112  \\
7     &      "     &     "     & 2006-03-03  &  3000   &    5        &  110  \\
8     & 12:56:21.2 & -76:37:47 & 2006-03-02  &  3000   &    2+2      &  114  \\
8     &      "     &     "     & 2006-03-03  &  3900   &    2+2      &  111  \\
9     & 12:57:42.7 & -77:25:31 & 2006-03-02  &  3000   &    0        &   80  \\
\tableline
10    & 13:04:00.1 & -76:49:59 & 2007-02-20  &  2700   &    0        &  127  \\
11    & 13:00:57.7 & -77:05:39 & 2007-02-20  &  2700   &    3        &  101  \\
12    & 13:04:44.3 & -77:47:03 & 2007-02-20  &  3600   &    3        &  107  \\
13    & 13:00:57.7 & -77:05:39 & 2007-02-20  &  2700   &    0        &  124  \\
14    & 13:10:01.8 & -77:00:56 & 2007-02-20  &  3600   &    1        &  112  \\
15    & 13:01:00.1 & -77:20:00 & 2007-02-21  &  3180   &    2        &   74  \\
\tableline  
\end{tabular}
\end{center}

\tablenotetext{\dagger}{For each pointing the coordinates of 
the MOS center are reported.}
\tablenotetext{\ddagger}{GM = GIRAFFE-MEDUSA, U = UVES. 
The LR06 grating was used for the GIRAFFE-MEDUSA configuration.}

\end{table}
\clearpage

\begin{table}
\begin{center}

\caption{Cha~II members candidates proposed in the literature,
but rejected by our spectroscopic criteria. \label{tab:rej}}

\begin{tabular}{lccl}

\tableline  
Object id.  &   RA       & DEC        &  Ref. \\        
\tableline  
\tableline  
C~30                   & 12:55:15.84 & $-$76:56:33.00 & \citet{Vuo01} \\
2MASS12560549-7654106  & 12:56:05.52 & $-$76:54:10.80 & \citet{You05} \\
C~37                   & 12:57:31.44 & $-$76:43:04.44 & \citet{Vuo01} \\
ISO-CHA\,II\,32        & 12:59:15.84 & $-$76:54:45.72 & \citet{Per03} \\
WFIJ12591881-7704419   & 12:59:18.81 & $-$77:04:41.92 & \citet{Spe07} \\
WFIJ12592348-7726589   & 12:59:23.48 & $-$77:26:58.96 & \citet{Spe07} \\
ISO-CHA\,II\,73        & 13:01:46.08 & $-$77:16:03.00 & \citet{Per03} \\
WFIJ13014752-7631023   & 13:01:47.52 & $-$76:31:02.32 & \citet{Spe07} \\
IRAS~12583-7634c$^\dag$& 13:02:05.52 & $-$76:51:02.88 & \citet{Sch91} \\
IRAS~12583-7634a$^\dag$& 13:02:08.88 & $-$76:51:06.12 & \citet{Sch91} \\
ISO-CHA\,II\,91        & 13:02:55.44 & $-$77:15:15.84 & \citet{Per03} \\
ChaII~376              & 13:03:12.48 & $-$76:50:50.64 & \citet{Lop05} \\
ISO-CHA\,II\,110       & 13:04:18.96 & $-$76:53:59.96 & \citet{Per03} \\
C~55                   & 13:05:32.64 & $-$77:35:26.16 & \citet{Vuo01} \\
C~45                   & 13:06:30.72 & $-$77:00:23.76 & \citet{Vuo01} \\
WFIJ13071960-7655476   & 13:07:19.60 & $-$76:55:47.64 & \citet{Spe07} \\
2MASS13102531-7729085  & 13:10:25.20 & $-$77:29:08.52 & \citet{You05} \\
\tableline
\end{tabular}
\end{center}
 \tablenotetext{\dag}{IRAS~12583-7634 was reported by \citet{Sch91} to be detected 
 only at 100$\mu$m. It was then associated to the ROSAT 
 source CHIIXR~12 by \citet{Alc00}. The object was not observed with IRAC 
 and, though falling in the area mapped with MIPS, it is not detected at 24~$\mu$m.
 Hence, it could not be investigated with the criteria by \citet{You05}.
 Within 10~arcsec from the position reported by \citet{Sch91} and \citet{Alc00},
 three stars are revealed in optical images, but none of them qualifies as PMS 
 from optical criteria \citep{Spe07}. Two of these stars were observed with FLAMES, 
 but none is a PMS star. The third one, i.e. IRAS~12583-7634b (RA=13:02:08.93 
 DEC=$-$76:51:05.72) could not be observed with FLAMES because of problems of fiber crowding.}     
\end{table}
\clearpage


\begin{deluxetable}{llllll}
\tabletypesize{\tiny}

\tablecaption{\label{pms_objects_b} Physical parameters for pre-main 
sequence objects and candidates in Cha~II. }

\tablewidth{0pt}

\rotate

\tablehead{
\colhead{No.} & \colhead{Object} & \colhead{T$_{\rm eff}$} & 
\colhead{A$_{\rm V}$} &  \colhead{log(L$^{\star}$/L$_{\odot}$)} & 
\colhead{R$^{\star}$/R$_{\odot}$}  \\ 
 & \colhead{id.} & [K] & [Mag] &   &   
}

\startdata

1   &   IRAS~12416-7703 / AX~Cha                   &  3487$\pm$200    &    3.36$\pm$1.50 &   1.11$\pm$0.15$^\spadesuit$&  9.75$\pm$0.64$^\spadesuit$	 \\  
2   &   IRAS~12448-7650                            &  3755$\pm$200    &    4.45$\pm$1.50 &   1.40$\pm$0.22$^\spadesuit$& 11.58$\pm$0.77$^\spadesuit$	 \\  
3   &   IRAS~F12488-7658 / C~13                    &  3025$\pm$200    &    4.32$\pm$1.50 &   0.43$\pm$0.17$^\spadesuit$&  6.09$\pm$0.31$^\spadesuit$	 \\  
4   &   IRAS~12496-7650 / DK~Cha                   &  7200$\pm$170    &   10.55$\pm$0.07 &   1.27$\pm$0.04	       &    2.77$\pm$0.14		 \\  
5   &   WFI~J12533662-7706393                      &  3000$\pm$200    &    5.58$\pm$1.50 &  -1.95$\pm$0.10	       &    0.39$\pm$0.02		 \\  
6   &   C~17                                       &  3600$\pm$200    &    5.95$\pm$1.50 &  -0.34$\pm$0.10	       &    1.74$\pm$0.09		 \\  
7   &   IRAS~12500-7658$^\bullet$                  &  2900$\pm$200    &    4.06$\pm$1.50 &  -2.21$\pm$0.10	       &    0.31$\pm$0.02		 \\  
8   &   C~33                                       &  3700$\pm$200    &    4.48$\pm$1.50 &  -0.78$\pm$0.24	       &    0.99$\pm$0.05		 \\  
9   &   IRAS~12522-7640  / HH~54                   &  --	      &        --	 &    --		       &	--			 \\  
10  &   Sz~46N                                     &  3705$\pm$72     &    2.44$\pm$0.30 &  -0.48$\pm$0.05	       &    1.39$\pm$0.09		 \\  
11  &   Sz~47                                      &  --	      &    --		 &    --		       &    --  			 \\  
12  &   IRAS~12535-7623 / CHIIXR~2                 &  3850$\pm$89     &    3.36$\pm$0.21 &   0.14$\pm$0.05	       &    2.71$\pm$0.16		 \\	
13  &   SSTc2d~J125758.7-770120 / \#1              &  2400$\pm$200    &    5.00$\pm$1.00 &  -3.05$\pm$0.10	       &    0.17$\pm$0.02		 \\  
14  &   ISO-CHAII 13$^\Omega$                      &  2880$\pm$70     &    5.00$\pm$0.50 &  -2.04$\pm$0.07	       &    0.38$\pm$0.03		 \\  
15  &   WFI~J12583675-7704065                      &  2400$\pm$200    &    3.84$\pm$1.50 &  -2.15$\pm$0.05	       &    0.49$\pm$0.03		 \\  
16  &   WFI~J12585611-7630105                      &  3025$\pm$70     &    1.66$\pm$0.98 &  -1.03$\pm$0.10	       &    1.13$\pm$0.13		 \\  
17  &   IRAS~12553-7651 / ISO-CHAII 28$^\bullet$   &  4500	      &    38.90	 &   1.20		       &    5.58			 \\  
18  &   C~41$^\Delta$                              &  3057$\pm$68     &    2.17$\pm$1.08 &  -1.95$\pm$0.11	       &    0.37$\pm$0.05		 \\  
19  &   ISO-CHAII 29                               &  3850$\pm$89     &    5.57$\pm$0.21 &  -0.19$\pm$0.05	       &    1.85$\pm$0.11		 \\  
20  &   IRAS~12556-7731                            &  3125$\pm$70     &    3.88$\pm$0.47 &   1.08$\pm$0.08$^\spadesuit$&   12.03$\pm$1.07$^\spadesuit$   \\  
21  &   WFI~J13005297-7709478                      &  2500$\pm$200    &    5.50$\pm$1.50 &  -2.08$\pm$0.29	       &    0.49$\pm$0.03		 \\  
22  &   Sz~48NE / CHIIXR~7                         &  3777$\pm$72     &    4.26$\pm$0.40 &  -0.47$\pm$0.06	       &    1.35$\pm$0.09		 \\  
23  &   Sz~49 / ISO-CHAII 55 / CHIIXR~9            &  3777$\pm$72     &    2.28$\pm$0.40 &  -0.70$\pm$0.06	       &    1.03$\pm$0.07		 \\  
24  &   Sz~48SW / CHIIXR~7                         &  3705$\pm$72     &    3.87$\pm$0.30 &  -0.58$\pm$0.05	       &    1.25$\pm$0.08		 \\  
25  &   Sz~50 / ISO-CHAII 52 / CHIIXR~8            &  3415$\pm$72     &    3.78$\pm$0.30 &   0.06$\pm$0.05	       &    3.10$\pm$0.19		 \\  
26  &   WFI~J13005531-7708295                      &  3687$\pm$72     &    2.72$\pm$0.24 &  -0.45$\pm$0.05	       &    1.46$\pm$0.09		 \\  
27  &   RXJ1301.0-7654a                            &  4350$\pm$192    &    1.93$\pm$0.47 &   0.38$\pm$0.06	       &    2.67$\pm$0.20		 \\  
28  &   IRAS~F12571-7657 / ISO-CHAII 54$^\dag$     &  --              &    --            &  --  		       &    --  			 \\  
29  &   Sz~51                                      &  3955$\pm$105    &    1.54$\pm$0.11 &  -0.36$\pm$0.05	       &    1.37$\pm$0.07		 \\  
30  &   CM~Cha / IRAS~12584-7621                   &  4060$\pm$125    &    1.52$\pm$0.14 &  -0.14$\pm$0.05	       &    1.78$\pm$0.10		 \\  
31  &   C~50                                       &  3125$\pm$70     &    2.95$\pm$0.98 &  -1.19$\pm$0.10	       &    0.89$\pm$0.10		 \\  
32  &   IRAS~12589-7646 / ISO-CHAII 89             &  3300$\pm$200    &    3.98$\pm$1.50 &   1.17$\pm$0.06$^\spadesuit$&   11.78$\pm$0.78$^\spadesuit$   \\  
33  &   RXJ1303.1-7706                             &  3850$\pm$89     &    1.61$\pm$0.21 &   0.10$\pm$0.05	       &    2.61$\pm$0.15		 \\  
34  &   C~51                                       &  3197$\pm$72     &    4.42$\pm$0.92 &  -0.53$\pm$0.09	       &    1.77$\pm$0.19		 \\  
35  &   WFI~J13031615-7629381                      &  2900$\pm$200    &    0.61$\pm$1.50 &  -1.24$\pm$0.22	       &    0.95$\pm$0.05		 \\  
36  &   Hn~22 / IRAS~13005-7633                    &  3560$\pm$72     &    0.61$\pm$0.24 &  -0.63$\pm$0.05	       &    1.24$\pm$0.07		 \\  
37  &   Hn~23                                      &  4350$\pm$192    &    1.24$\pm$0.47 &  -0.06$\pm$0.04	       &    1.60$\pm$0.14		 \\    
38  &   Sz~52                                      &  3487$\pm$72     &    4.14$\pm$0.24 &  -0.75$\pm$0.05	       &    1.15$\pm$0.07		 \\    
39  &   Hn~24                                      &  3850$\pm$89     &    2.76$\pm$0.38 &   0.02$\pm$0.06	       &    2.37$\pm$0.16		 \\	
40  &   Hn~25                                      &  3487$\pm$72     &    4.10$\pm$0.24 &  -0.48$\pm$0.05	       &    1.56$\pm$0.09		 \\	
41  &   Sz~53                                      &  3705$\pm$72     &    3.68$\pm$0.31 &  -0.49$\pm$0.05	       &    1.39$\pm$0.09		 \\  
42  &   Sz~54                                      &  4350$\pm$192    &    1.57$\pm$0.47 &   0.29$\pm$0.07	       &    2.42$\pm$0.20		 \\  
43  &   SSTc2d~J130521.7-773810$^\dag$             &  --              &    --            &    --		       &    --  			 \\   
44  &   SSTc2d~J130529.0-774140                    &  --	      &    --		 &    --		       &    --  			 \\  
45  &   SSTc2d~J130540.8-773958 / \#5              &  2200$\pm$300    &    3.00$\pm$1.00 &  -3.22$\pm$0.14	       &    0.17$\pm$0.02		 \\  
46  &   Sz~55                                      &  3560$\pm$72     &    3.09$\pm$0.22 &  -0.90$\pm$0.06	       &    0.91$\pm$0.06		 \\  
47  &   Sz~56                                      &  3270$\pm$72     &    3.18$\pm$0.64 &  -0.47$\pm$0.08	       &    1.78$\pm$0.17		 \\  
48  &   Sz~57 / C~60                               &  3125$\pm$70     &    3.09$\pm$0.98 &  -0.39$\pm$0.10	       &    2.21$\pm$0.26		 \\  
49  &   Sz~58 / IRAS~13030-7707 /C~61              &  4350$\pm$192    &    3.87$\pm$0.47 &  -0.16$\pm$0.07	       &    1.43$\pm$0.12		 \\  
50  &   Sz~59                       		   &  4060$\pm$125    &    2.67$\pm$0.16 &  -0.05$\pm$0.03	       &    1.96$\pm$0.12		 \\  
51  &   C~62                       		   &  3197$\pm$72     &    4.84$\pm$0.92 &  -1.05$\pm$0.10	       &    0.97$\pm$0.11		 \\  
52  &   Sz~60W                    		   &  3705$\pm$72     &    2.35$\pm$0.32 &  -0.54$\pm$0.06	       &    1.32$\pm$0.10		 \\  
53  &   Sz~60E                    		   &  3270$\pm$72     &    2.16$\pm$0.64 &  -0.67$\pm$0.08	       &    1.42$\pm$0.13		 \\  
54  &   IRAS~13036-7644 / BHR~86                   &  --	      &    --		 &    --		       &    --  			 \\  
55  &   Hn~26                                      &  3560$\pm$72     &    3.50$\pm$0.22 &  -0.59$\pm$0.06	       &    1.30$\pm$0.08		 \\   
56  &   Sz~61                                      &  4350$\pm$192    &    3.13$\pm$0.47 &   0.07$\pm$0.05	       &    1.87$\pm$0.15		 \\  
57  &   C~66                                       &  3197$\pm$72     &    4.37$\pm$0.92 &  -1.30$\pm$0.10	       &    0.73$\pm$0.08		 \\   
58  &   IRAS~F13052-7653NW / CHIIXR~60             &  3777$\pm$72     &    2.28$\pm$0.40 &  -0.70$\pm$0.07	       &    1.03$\pm$0.08		 \\   
59  &   IRAS~F13052-7653N  / CHIIXR~60             &  3632$\pm$72     &    0.41$\pm$0.20 &  -0.47$\pm$0.06	       &    1.49$\pm$0.10		 \\   
60  &   Sz~62                                      &  3487$\pm$72     &    0.78$\pm$0.24 &  -0.48$\pm$0.06	       &    1.56$\pm$0.10		 \\  
61  &   Sz~63                                      &  3415$\pm$72     &    1.61$\pm$0.30 &  -0.64$\pm$0.06	       &    1.38$\pm$0.10		 \\  
62  &   2MASS13125238-7739182                      &  3197$\pm$72     &    1.34$\pm$0.93 &  -0.81$\pm$0.10	       &    1.28$\pm$0.15		 \\	  
63  &   Sz~64                                      &  3125$\pm$70     &    0.74$\pm$0.50 &  -1.04$\pm$0.21	       &    1.04$\pm$0.13		 \\  

\enddata

\tablecomments{The coordinates are reported in Paper~I}
\tablenotetext{\spadesuit}{Estimates by assuming the Cha~II distance of 178~pc (see also \S~\ref{sec_hrd})}
\tablenotetext{\bullet}{Physical parameters have been estimated as explained in See \S~\ref{classI}}
\tablenotetext{\dag}{Object classified as ``Continuum-type'' (see \S~\ref{SpecTyp})}
\end{deluxetable}


\clearpage


\begin{deluxetable}{ll|cc|cc|cc}
\tabletypesize{\tiny}

\tablecaption{\label{tab:par2} Masses and ages for the pre-main 
sequence objects and candidates in Cha~II.}

\tablewidth{0pt}

\rotate

\tablehead{

& & B98\&C00$^\dag$ & & DM98$^\ddag$  & & PS99$^\aleph$   \\

\colhead{No.} & \colhead{Object id.} & \colhead{M$_\star$/M$_\odot$} & \colhead{Age} & \colhead{M$_\star$/M$_\odot$} &  
\colhead{Age} & \colhead{M$_\star$/M$_\odot$} &   \colhead{Age} \\
 & & & (Myr) &  & (Myr) & & (Myr)
}

\startdata

1   &  IRAS~12416-7703 / AX~Cha$^\spadesuit$ 	   &	--   &      --  &    --  &    --   &   --   &    --    \\  
2   &  IRAS~12448-7650$^\spadesuit$ 		   &	--   &      --  &    --  &    --   &   --   &    --    \\  
3   &  IRAS~F12488-7658 / C~13$^\spadesuit$  	   &	--   &      --  &    --  &    --   &   --   &    --    \\  
4   &  IRAS~12496-7650 / DK~Cha 		   &	--   &      --  &  2.00  &   7.0   & 2.00   &	2.5    \\  
5   &  WFI~J12533662-7706393			   &  0.07   &    7.2   &  0.10  &  10.0   & 0.10   &  10.0    \\  
6   &  C~17					   &  0.62   &    2.2   &  0.30  &   0.7   & 0.40   &	1.0    \\  
7   &  IRAS~12500-7658$^\bullet$		   &  0.06   &   10.1   &  0.06  &  10.0   &   --   &	 --    \\  
8   &  C~33					   &  0.70   &   14.3   &  0.50  &   7.0   & 0.50   &	7.5    \\  
9   &  IRAS~12522-7640  / HH~54 		   &	--   &      --  &    --  &     --  &   --   &    --    \\  
10  &  Sz~46N					   &  0.7    &    5.7   &  0.40  &   1.5   & 0.50   &	2.0    \\  
11  &  Sz~47					   &	--   &      --  &    --  &     --  &	--  &    --    \\  
12  &  IRAS~12535-7623 / CHIIXR~2		   &  1.05   &    1.1   &  0.35  &   0.3   & 0.60   &	1.0    \\  
13  &  SSTc2d~J125758.7-770120 / \#1		   &  0.03   &   28.6   &  0.02  &   7.0   &   --   &	 --    \\  
14  &  ISO-CHAII 13				   &  0.05   &    5.0   &  0.06  &   7.0   &   --   &	 --    \\  
15  &  WFI~J12583675-7704065			   &  0.02   &    1.0   &  0.08  &   0.2   &   --   &    --    \\  
16  &  WFI~J12585611-7630105			   &  0.11   &    1.0   &  0.14  &   2.0   & 0.10   &   1.0    \\  
17  &  IRAS~12553-7651 / ISO-CHAII 28$^\bullet$    &	--   &     --   &    --  &    --   &   --   &    --    \\  
18  &  C~41					   &  0.09   &   10.1   &  0.12  &  20.0   & 0.10   &  10.0    \\  
19  &  ISO-CHAII 29				   &  0.95   &    5.1   &  0.40  &   0.7   & 0.60   &   2.5    \\  
20  &  IRAS~12556-7731$^\spadesuit$  		   &	--   &      --  &    --  &    --   &   --   &    --    \\  
21  &  WFI~J13005297-7709478			   &  0.02   &    1.0   &  0.03  &   0.1   &   --   &    --    \\  
22  &  Sz~48NE / CHIIXR~7			   &  0.75   &    8.0   &  0.50  &   3.0   & 0.50   &	4.0    \\  
23  &  Sz~49 / ISO-CHAII 55 / CHIIXR~9  	   &  0.75   &   14.3   &  0.60  &   7.0   & 0.50   &	7.5    \\  
24  &  Sz~48SW / CHIIXR~7			   &  0.70   &    7.1   &  0.50  &   3.0   & 0.50   &	4.0    \\  
25  &  Sz~50 / ISO-CHAII 52 / CHIIXR~8  	   &  0.50   &    1.0   &  0.25  &   0.1   & 0.30   &   0.0    \\  
26  &  WFI~J13005531-7708295			   &  0.70   &    4.5   &  0.40  &   2.0   & 0.50   &	2.0    \\  
27  &  RXJ1301.0-7654a  			   &	--   &     --   &  0.50  &   0.2   & 0.90   &	1.0    \\  
28  &  IRAS~F12571-7657 / ISO-CHAII 54$^\Delta$    &    --   &     --   &  --    &   --    &  --    &	--     \\  
29  &  Sz~51					   &  1.00   &    9.0   &  0.50  &   1.0   & 0.70   &	4.0    \\  
30  &  CM~Cha / IRAS~12584-7621 		   &  1.15   &    5.0   &  0.50  &   1.0   & 0.90   &	5.0    \\  
31  &  C~50					   &  0.18   &    3.2   &  0.18  &   3.0   & 0.15   &	2.0    \\  
32  &  IRAS~12589-7646 / ISO-CHAII 89$^\spadesuit$ &	--   &      --  &    --  &    --   &   --   &    --    \\  
33  &  RXJ1303.1-7706				   &  1.05   &    1.4   &  0.35  &   0.3   & 0.80   &	1.5    \\  
34  &  C~51					   &  0.25   &    1.0   &  0.16  &   0.2   & 0.15   &	0.7    \\  
35  &  WFI~J13031615-7629381			   &  0.07   &    1.0   &  0.12  &   3.0   & 0.10   &	2.5    \\  
36  &  Hn~22 / IRAS~13005-7633  		   &  0.50   &    4.0   &  0.35  &   2.0   & 0.40   &	2.5    \\  
37  &  Hn~23					   &  1.40   &    5.7   &  0.70  &   1.0   & 0.90   &	4.0    \\  
38  &  Sz~52					   &  0.45   &    4.5   &  0.30  &   2.0   & 0.30   &	2.0    \\  
39  &  Hn~24					   &  1.00   &    1.6   &  0.35  &   0.5   & 0.60   &	1.0    \\  
40  &  Hn~25					   &  0.50   &    2.3   &  0.30  &   1.0   & 0.30   &	1.5    \\  
41  &  Sz~53					   &  0.75   &    5.7   &  0.40  &   2.0   & 0.50   &	2.0    \\  
42  &  Sz~54					   &  1.40   &    1.6   &  0.60  &   0.5   & 0.90   &	2.0    \\  
43  &  SSTc2d~J130521.7-773810$^\Delta$  	   &	 --  &      --  &   --   &    --   &  --    &    --    \\  
44  &  SSTc2d~J130529.0-774140  		   &	 --  &      --  &   --   &    --   &  --    &    --    \\  
45  &  SSTc2d~J130540.8-773958 / \#5		   &  0.02   &   25.2   &  0.02  &  10.0   &  --    &    --    \\  
46  &  Sz~55					   &  0.50   &   10.0   &  0.35  &   5.0   & 0.40   &	7.5    \\  
47  &  Sz~56					   &  0.30   &    1.0   &  0.18  &   0.7   & 0.20   &	0.7    \\  
48  &  Sz~57 / C~60				   &  0.15   &    1.0   &  0.14  &   0.1   & 0.15   &   0.7    \\  
49  &  Sz~58 / IRAS~13030-7707 /C~61		   &  1.20   &   11.3   &  0.70  &   2.0   & 1.00   &	7.5    \\  
50  &  Sz~59					   &  1.15   &    3.6   &  0.50  &   0.5   & 0.80   &	2.5    \\  
51  &  C~62					   &  0.20   &    2.5   &  0.18  &   2.0   & 0.15   &	1.0    \\  
52  &  Sz~60W					   &  0.70   &    5.7   &  0.50  &   3.0   & 0.50   &	2.0    \\  
53  &  Sz~60E					   &  0.30   &    1.6   &  0.18  &   1.0   & 0.20   &	1.0    \\  
54  &  IRAS~13036-7644 / BHR~86 		   &   --    &      --  &   --   &    --   &   --   &    --    \\  
55  &  Hn~26					   &  0.57   &    4.5   &  0.35  &   2.0   & 0.40   &	2.5    \\  
56  &  Sz~61					   &  1.40   &    4.5   &  0.60  &   1.0   & 1.00   &	2.5    \\  
57  &  C~66					   &  0.17   &    4.5   &  0.20  &   5.0   & 0.15   &  10.0    \\  
58  &  IRAS~F13052-7653NW / CHIIXR~60		   &  0.75   &   14.3   &  0.60  &   7.0   & 0.50   &	2.0    \\  
59  &  IRAS~F13052-7653N  / CHIIXR~60		   &  0.62   &    3.6   &  0.40  &   2.0   & 0.40   &	2.5    \\  
60  &  Sz~62					   &  0.50   &    2.3   &  0.30  &   1.0   & 0.30   &	1.5    \\  
61  &  Sz~63					   &  0.40   &    2.5   &  0.25  &   1.0   & 0.30   &	1.0    \\  
62  &  2MASS13125238-7739182			   &  0.25   &    1.3   &  0.18  &   1.5   & 0.15   &	1.5    \\  
63  &  Sz~64					   &  0.18   &    2.0   &  0.16  &   2.0   & 0.15   &   1.0    \\  

\enddata

\tablenotetext{\dag}{B98\&C00 = \citet{Bar98} \& \citet{Cha00}; models available for 0.003$<M_\star<$1.4~M$_{\odot}$}
\tablenotetext{\ddag}{DM98 = \citet{DAn98}; models available for 0.01$<M_\star<$3~M$_{\odot}$}
\tablenotetext{\aleph}{PS99 = \citet{Pal99}; models available for 0.1$<M_\star<$6~M$_{\odot}$}
\tablenotetext{\bullet}{See \S~\ref{classI}}
\tablenotetext{\spadesuit}{Mass and age could not be estimated (see \S~\ref{sec_hrd}).}
\tablenotetext{\Delta}{Object classified as ``Continuum-type'' (see \S~\ref{SpecTyp})}

\end{deluxetable}



\begin{deluxetable}{c|l|c|c|c}
\tabletypesize{\scriptsize}

\tablecaption{\label{tab:SF} Overall results on the star formation in Cha~II.}
\rotate

\tablewidth{0pt}

\tablehead{
       & &  B98\&C00$^{\dag}$ &  DM98$^{\ddag}$ &  PS99$^{\aleph}$ \\  
}

\startdata

     & Mean mass (M$_{\odot}$)            & 0.59           & 0.38          & 0.50          \\
     & Total mass$^\Theta$ (M$_{\odot}$)  & 30.0           & 22.0          & 24.6          \\
PMS  & IMF slope$^\Delta$ ($\alpha$)      & 0.36$\pm$0.19  & 0.84$\pm$0.30 & 0.37$\pm$0.20 \\
stars& Mean age (Myr)                     & 5.76           & 2.87          & 2.83          \\
     & SFE$^\Omega$ (\%)                  & 2.3-4.1        & 1.7-3.0       & 1.9-3.4       \\
     & SF~rate (M$_{\odot}$/Myr)          & 5.4            & 7.6           & 8.69          \\

\tableline
             & Mean mass (M$_{\odot}$)           & 0.55          & 0.36          & 0.48           \\
             & Total mass$^\Theta$ (M$_{\odot}$) & 31.4          & 23.0	         & 25.8	          \\
PMS stars    & IMF slope$^\Delta$ ($\alpha$)     & 0.48$\pm$0.22 & 0.92$\pm$0.28 & 0.42$\pm$0.19  \\
+candidates  & Mean age (Myr)                    & 5.60          & 2.94          & 3.05           \\
             & SFE$^\Omega$ (\%)                 & 2.5-4.3       & 1.8-3.2       & 2.0-3.5        \\
             & SF~rate (M$_{\odot}$/Myr)         & 5.6  	 & 7.8  	 & 8.5 	          \\

\tableline
     & Mean mass (M$_{\odot}$)           &   0.62    & 0.34    & 0.47	  \\
     & Total mass$^\Theta$ (M$_{\odot}$) &   15.0    & 10.6    & 11.0	  \\
YSOs & Mean age (Myr)                    &   4.06    & 2.22    & 2.50	  \\
     & SFE$^\Omega$ (\%)                 & 1.2-2.1   & 0.8-1.5 & 0.9-1.5  \\
     & SF~rate (M$_{\odot}$/Myr)         &   3.7     & 4.8     & 4.4	  \\

\enddata
\tablenotetext{\dag}{B98\&C00 = \citet{Bar98} \& \citet{Cha00}}
\tablenotetext{\ddag}{DM98 = \citet{DAn98}}
\tablenotetext{\aleph}{PS99 = \citet{Pal99}}
\tablenotetext{\Theta}{For the 8 object whose mass could not be estimated the mean mass 
of the Cha~II members was assumed (see \S~\ref{sec_hrd} and \S~\ref{SFE})}
\tablenotetext{\Delta}{Mass range 0.1$ < M < $1~M$_{\odot}$}
\tablenotetext{\Omega}{The uncertainty on SFE depends on
the uncertainty on cloud mass (see \S~\ref{SFE})}

\end{deluxetable}


\end{document}